\documentclass[journal]{IEEEtran}
\usepackage{cite}
\usepackage{amsmath}
\usepackage{algorithmic}
\usepackage{array}

\ifCLASSOPTIONcompsoc
\usepackage[caption=false,font=normalsize,labelfont=sf,textfont=sf]{subfig}
\else
\usepackage[caption=false,font=footnotesize]{subfig}
\fi

\usepackage{dblfloatfix}

\ifCLASSOPTIONcaptionsoff
\usepackage[nomarkers]{endfloat}
\let\MYoriglatexcaption\caption
\renewcommand{\caption}[2][\relax]{\MYoriglatexcaption[#2]{#2}}
\fi

\usepackage{url}

\usepackage{amssymb,amsfonts,mathtools}
\usepackage{amsthm,amscd,amsbsy}
\usepackage[ruled]{algorithm2e}

\usepackage{float}

\usepackage{graphicx,epsfig,color,graphics,caption}
\usepackage{epsf,psfrag,hhline,textcomp}
\usepackage{xcolor}
\usepackage{times,fancyhdr}
\usepackage{multirow}
\usepackage{latexsym}
\usepackage{makecell}
\usepackage{comment}
\usepackage{tabularx} 
\usepackage{paralist}
\usepackage{booktabs}
\usepackage{enumerate}
\usepackage{hyperref} 
\hypersetup{
	bookmarks=true,
	unicode=true,
	pdftoolbar=true,
	pdfmenubar=true,
	pdftitle={A Time-dependent SIR model for COVID-19 with Undetectable Infected Persons},
	pdfauthor={Yi-Cheng Chen, Ping-En Lu, Cheng-Shang Chang, Tzu-Hsuan Liu},
	pdfsubject={Populations and Evolution (q-bio.PE), Machine Learning (cs.LG), Machine Learning (stat.ML)},
	pdfcreator={Ping-En Lu},
	pdfkeywords={COVID-19, SARS-CoV-2, 2019-nCoV, Coronavirus, Time-dependent SIR model, SIR, asymptomatic infection, herd immunity, superspreader, independent cascade, social distancing},
	pdfnewwindow=true,
	colorlinks=true,
	linkcolor=blue
}

\usepackage[utf8]{inputenc}
\usepackage[english]{babel}
\usepackage[export]{adjustbox}
\usepackage{sidecap}
\usepackage{environ}
\newcommand{\mymarginpar}[1]{\marginpar{#1}}
\renewcommand{\marginpar}[1]{}

\newcommand{\bearn}{\begin{eqnarray*}}
	\newcommand{\eearn}{\end{eqnarray*}}
\newcommand{\barr}{\begin{array}}
	\newcommand{\earr}{\end{array}}

\newcommand{\N}{{\cal N}}

\newtheorem{definition}{Definition}
\newtheorem{assumption}[definition]{Assumption}
\newtheorem{property}[definition]{Property}
\newtheorem{proposition}[definition]{Proposition}
\newtheorem{lemma}[definition]{Lemma}
\newtheorem{theorem}[definition]{Theorem}
\newtheorem{corollary}[definition]{Corollary}
\newtheorem{example}[definition]{Example}
\newtheorem{remark}[definition]{Remark}
\newtheorem{axiom}[definition]{Axiom}

\newcommand{\benum}{\begin{enumerate}}
	\newcommand{\eenum}{\end{enumerate}}
\newcommand{\bdesc}{\begin{description}}
	\newcommand{\edesc}{\end{description}}

\newcommand{\bdefin}[1]{\begin{definition}\mymarginpar{def:#1}\label{def:#1}}
	\newcommand{\edefin}{\end{definition}}

\newcommand{\bpro}[1]{\begin{property}\mymarginpar{pro:#1}\label{pro:#1}}
	\newcommand{\epro}{\end{property}}

\newcommand{\bprop}[1]{\begin{proposition}\mymarginpar{prop:#1}\label{prop:#1}}
	\newcommand{\eprop}{\end{proposition}}

\newcommand{\blem}[1]{\begin{lemma}\mymarginpar{lem:#1}\label{lem:#1}}
	\newcommand{\elem}{\end{lemma}}

\newcommand{\bass}[1]{\begin{assumption}\mymarginpar{the:#1}\label{ass:#1}}
	\newcommand{\eass}{\end{assumption}}

\newcommand{\bthe}[1]{\begin{theorem}\mymarginpar{the:#1}\label{the:#1}}
	\newcommand{\ethe}{\end{theorem}}
\newcommand{\rthe}[1]{Theorem \ref{the:#1}}
\newcommand{\bproof}{\noindent{\bf Proof.}}
\newcommand{\eproof}{\hfill \squares \\ \vspace{.3cm}}
\newcommand{\bcor}[1]{\begin{corollary}\mymarginpar{cor:#1}\label{cor:#1}}
	\newcommand{\ecor}{\end{corollary}}
\newcommand{\rcor}[1]{Corollary \ref{cor:#1}}
\newcommand{\bax}[1]{\begin{axiom}\mymarginpar{ax:#1}\label{ax:#1}}
	\newcommand{\eax}{\vspace{-.1in} \end{axiom}}

\newcommand{\bex}[2]{\vspace{.1in}\begin{example}\mymarginpar{ex:#1}{\bf #2}\label{ex:#1}}
	\newcommand{\eex}{\end{example}\vspace{.3cm}}

\newcommand{\brem}[1]{\begin{remark}\mymarginpar{rem:#1}\label{rem:#1}\em}
	\newcommand{\erem}{\end{remark}}

\newcommand{\beq}[1]{\mymarginpar{eq:#1}\begin{equation}\label{eq:#1}}
\newcommand{\beqno}[1]{\mymarginpar{eq:#1}\begin{eqnarray}\nonumber}
\newcommand{\eeq}{\end{equation}}
\newcommand{\eeqno}{&&\end{eqnarray}}
\newcommand{\req}[1]{(\ref{eq:#1})}

\newcommand{\bear}[1]{\mymarginpar{eq:#1}\begin{eqnarray}\label{eq:#1}}
\newcommand{\bearno}[1]{\mymarginpar{eq:#1}\begin{eqnarray}\nonumber}
\newcommand{\eear}{\end{eqnarray}}
\newcommand{\eearno}{\end{eqnarray}}
\NewEnviron{es}{
\begin{equation}\begin{split}
\BODY
\end{split}\end{equation}
}
\newcommand{\bieeeeq}[1]{\mymarginpar{eq:#1}\begin{IEEEeqnarray}{rCl}\label{eq:#1}}
\newcommand{\eieeeeq}{\end{IEEEeqnarray}}
\newcommand{\bsel}{\left \{\begin{array}{cl}}
\newcommand{\esel}{\end{array}\right .}
\newcommand{\bmat}[1]{\left [\begin{array}{#1}}
\newcommand{\emat}{\end{array}\right ]}

\newcommand{\bsubsec}[2]{\mymarginpar{subsec:#2}\subsection{#1}\label{subsec:#2}}
\newcommand{\rsubsec}[1]{Subsection \ref{subsec:#1}}

\newcommand{\bapp}{\begin{appendices}}
\newcommand{\eapp}{\end{appendices}}
\def\R{I\kern-0.30em R}
\def\N{I\kern-0.30em N}
\def\P{I\kern-0.30em P}

\newcommand \squares{\vrule height6pt width7pt depth1pt}

\newcommand{\rfig}[1]{Figure \ref{fig:#1}}

\newcommand{\RomanNumeralCaps}[1]
{\MakeUppercase{\romannumeral #1}}

\begin{document}
	
\title{A Time-dependent SIR model for COVID-19 with Undetectable Infected Persons}

\author{
	\IEEEauthorblockN{Yi-Cheng~Chen\IEEEauthorrefmark{1}, Ping-En~Lu\IEEEauthorrefmark{2},~\IEEEmembership{Graduate~Student~Member,~IEEE,} Cheng-Shang~Chang\IEEEauthorrefmark{3},~\IEEEmembership{Fellow,~IEEE,} and Tzu-Hsuan~Liu\IEEEauthorrefmark{4}}

	\IEEEauthorblockA{Institute of Communications Engineering\\
		National Tsing Hua University\\
		Hsinchu 30013, Taiwan, R.O.C.\\
		Email: \IEEEauthorrefmark{1}yichengchen@gapp.nthu.edu.tw, \IEEEauthorrefmark{2}j94223@gmail.com, \IEEEauthorrefmark{3}cschang@ee.nthu.edu.tw, \IEEEauthorrefmark{4}carinaliu@gapp.nthu.edu.tw}%
	}%

\markboth{The latest version will be placed on \url{http://gibbs1.ee.nthu.edu.tw/A_TIME_DEPENDENT_SIR_MODEL_FOR_COVID_19.PDF}}%
{Chen \MakeLowercase{\textit{et al.}}: A Time-dependent SIR model for COVID-19 with Undetectable Infected Persons}

\IEEEspecialpapernotice{The latest version will be placed on this link: \url{http://gibbs1.ee.nthu.edu.tw/A_TIME_DEPENDENT_SIR_MODEL_FOR_COVID_19.PDF}}

\maketitle

\begin{abstract}
In this paper, we conduct mathematical and numerical analyses to address the following important questions for COVID-19: (Q1) Is it possible to contain COVID-19? (Q2) If COVID-19 can be contained, when will be the peak of the epidemic, and when will it end? (Q3) How do the asymptomatic infections affect the spread of disease? (Q4) If COVID-19 cannot be contained, what is the ratio of the population that needs to be infected in order to achieve herd immunity? (Q5) How effective are the social distancing approaches? (Q6) If COVID-19 cannot be contained, what is the ratio of the population infected in the long run? For (Q1) and (Q2), we propose a time-dependent susceptible-infected-recovered (SIR) model that tracks two time series: (i) the transmission rate at time $t$ and (ii) the recovering rate at time $t$. Such an approach is not only more adaptive than traditional static SIR models, but also more robust than direct estimation methods. Using the data provided by the National Health Commission of the People's Republic of China (NHC) \cite{outbreak_notification_2020}, we show that the one-day prediction errors for the numbers of confirmed cases are almost less than $3\%$. Also, the turning point, defined as the day that the transmission rate is less than the recovering rate, is predicted to be Feb. 17, 2020. After that day, the basic reproduction number, known as the $R_0$ value at time $t$, is less than $1$. In that case, the total number of confirmed cases is predicted to be around $80,000$ cases in China under our model. For (Q3), we extend our SIR model by considering two types of infected persons: detectable infected persons and undetectable infected persons. Whether there is an outbreak in such a model is characterized by the spectral radius of a $2 \times 2$ matrix that is closely related to the basic reproduction number $R_0$. We plot the phase transition diagram of an outbreak and show that there are several countries, including South Korea, Italy, and Iran, that are on the verge of COVID-19 outbreaks on Mar. 2, 2020. For (Q4), we show that herd immunity can be achieved after at least $1-\frac{1}{R_0}$ fraction of individuals being infected and recovered from COVID-19. For (Q5) and (Q6), we analyze the independent cascade (IC) model for disease propagation in a random network specified by a degree distribution. By relating the propagation probabilities in the IC model to the transmission rates and recovering rates in the SIR model, we show two approaches of social distancing that can lead to a reduction of $R_0$.
\end{abstract}


\IEEEpeerreviewmaketitle

\section{Introduction}\label{intro}
At the beginning of December 2019, the first COVID-19 victim was diagnosed with the coronavirus in Wuhan, China. In the following weeks, the disease spread widely in China mainland and other countries, which causes global panic. The virus has been named “SARS-CoV-2,” and the disease it causes has been named “coronavirus disease 2019 (abbreviated “COVID-19”). There have been $80,151$ people infected by the disease and $2,943$ deaths until Mar. 2, 2020 according to the official statement by the Chinese government. To block the spread of the virus, there are some strategies such as city-wide lockdown, traffic halt, community management, social distancing, and propaganda of health education knowledge that have been adopted by the governments of China and other countries in the world.

Unlike the Severe Acute Respiratory Syndrome (SARS) and other infectious diseases, one problematic characteristic of COVID-19 is that there are asymptomatic infections (who have very mild symptoms). Those asymptomatic infections are unaware of their contagious ability, and thus get more people infected. The transmission rate can increase dramatically in this circumstance. According to the recent report from WHO \cite{who}, only $87.9\%$ of COVID-19 patients have a fever, and $67.7\%$ of them have a dry cough. If we use body temperature as a means to detect COVID-19 infected cases, then more than $10\%$ of infected persons cannot be detected.

Due to the recent development of the epidemic, we are interested in addressing the following important questions for COVID-19:
\begin{description}
	\item[(Q1)] Is it possible to contain COVID-19? Are the commonly used measures, such as city-wide lockdown, traffic halt, community management, and propaganda of health education knowledge, effective in containing COVID-19?
	\item[(Q2)] If COVID-19 can be contained, when will be the peak of the epidemic, and when will it end?
	\item[(Q3)] How do the asymptomatic infections affect the spread of disease?
	\item[(Q4)] If COVID-19 cannot be contained, what is the ratio of the population that needs to be infected in order to achieve herd immunity?
	\item[(Q5)] How effective are the social distancing approaches, such as reduction of interpersonal contacts and canceling mass gatherings in controlling COVID-19?
	\item[(Q6)] If COVID-19 cannot be contained, what is the ratio of the population infected in the long run?
\end{description}

For (Q1), we analyze the cases in China and aim to predict how the virus spreads in this paper. Specifically, we propose using a time-dependent susceptible-infected-recovered (SIR) model to analyze and predict the number of infected persons and the number of recovered persons (including deaths). In the traditional SIR model, it has two time-invariant variables: the transmission rate $\beta$ and the recovering rate $\gamma$. The transmission rate $\beta$ means that each individual has on average $\beta$ contacts with randomly chosen others per unit time. On the other hand, the recovering rate $\gamma$ indicates that individuals in the infected state get recovered or die at a fixed average rate $\gamma$. The traditional SIR model neglects the time-varying property of $\beta$ and $\gamma$, and it is too simple to precisely and effectively predict the trend of the disease. Therefore, we propose using a time-dependent SIR model, where both the transmission rate $\beta$ and the recovering rate $\gamma$ are functions of time $t$. Our idea is to use machine learning methods to track the transmission rate $\beta(t)$ and the recovering rate $\gamma(t)$, and then use them to predict the number of the infected persons and the number of recovered persons at a certain time $t$ in the future. Our time-dependent SIR model can dynamically adjust the crucial parameters, such as $\beta(t)$ and $\gamma(t)$, to adapt accordingly to the change of control policies, which differs from the existing SIR and SEIR models in the literature, e.g., \cite{nesteruk2020statistics}, \cite{chen2020time}, \cite{peng2020epidemic}, \cite{zhou2020preliminary}, and \cite{maier2020effective}. For example, we observe that city-wide lockdown can lower the transmission rate substantially from our model. Most data-driven and curve-fitting methods for the prediction of COVID-19, e.g., \cite{zhao2020preliminary}, \cite{zeng2020predictions}, and \cite{hu2020artificial} seem to track data perfectly; however, they are lack of physical insights of the spread of the disease. Moreover, they are very sensitive to the sudden change in the definition of confirmed cases on Feb. 12, 2020 in the Hubei province. On the other hand, our time-dependent SIR model can examine the epidemic control policy of the Chinese government and provide reasonable explanations. Using the data provided by the National Health Commission of the People's Republic of China (NHC) \cite{outbreak_notification_2020}, we show that the one-day prediction errors for the numbers of confirmed cases are almost less than $3\%$ except for Feb. 12, 2020, which is unpredictable due to the change of the definition of confirmed cases.

For (Q2), the basic reproduction number $R_0$, defined as the number of additional infections by an infected person before it recovers, is one of the commonly used metrics to check whether the disease will become an outbreak. In the classical SIR model, $R_0$ is simply $\beta/\gamma$ as an infected person takes (on average) $1/\gamma$ days to recover, and during that period time, it will be in contact with (on average) $\beta$ persons. In our time-dependent SIR model, the basic reproduction number $R_0(t)$ is a function of time, and it is defined as $\beta(t) / \gamma (t)$. If $R_0(t) > 1$, the disease will spread exponentially and infects a certain fraction of the total population. On the contrary, the disease will eventually be contained. Therefore, by observing the change of $R_0(t)$ with respect to time or even predict $R_0(t)$ in the future, we can check whether certain epidemic control policies are effective or not. Using the data provided by the National Health Commission of the People's Republic of China (NHC) \cite{outbreak_notification_2020}, we show that the turning point (peak), defined as the day that the basic reproduction number is less than $1$, is predicted to be Feb. 17, 2020. Moreover, the disease in China will end in about $6$ weeks after its peak in our (deterministic) model if the current contagious disease control policies are maintained in China. In that case, the total number of confirmed cases is predicted to be around $80,000$ cases in China under our (deterministic) model.

For (Q3), we extend our SIR model to include two types of infected persons: detectable infected persons (type \RomanNumeralCaps{1}) and undetectable infected persons (type \RomanNumeralCaps{2}). With probability $w_1$ (resp. $w_2$), an infected person is of type \RomanNumeralCaps{1} (resp. \RomanNumeralCaps{2}), where $w_1+w_2=1$. Type \RomanNumeralCaps{1} (resp. \RomanNumeralCaps{2}) infected persons have the transmission rate $\beta_1$ (resp. $\beta_2$) and the recovering rate $\gamma_1$ (resp. $\gamma_2$). The basic reproduction number in this model is
\beq{imp1111}
R_0 =w_1\frac{\beta_1}{\gamma_1}+w_2\frac{\beta_2}{\gamma_2}.
\eeq
In practice, type \RomanNumeralCaps{1} infected persons have a lower transmission rate than that of type \RomanNumeralCaps{2} infected persons (as type \RomanNumeralCaps{1} infected persons can be isolated). For such a model, whether the disease is controllable is characterized by the spectral radius of a $2 \times 2$ matrix. If the spectral radius of that matrix is larger than $1$, then there is an outbreak. On the other hand, if it is smaller than $1$, then there is no outbreak. One interesting result is that the spectral radius of that matrix is larger (resp. smaller) than $1$ if the basic reproduction number $R_0$ in \req{imp1111} is larger (resp. smaller) than $1$. The curve that has the spectral radius equal to $1$ is known as the percolation threshold curve in a phase transition diagram \cite{newman2010networks}. Using the historical data from Jan. 22, 2020 to Mar. 2, 2020 from the GitHub of Johns Hopkins University \cite{cssegisanddata_johns}, we extend our study to some other countries, including Japan, Singapore, South Korea, Italy, and Iran. Our numerical results show that there are several countries, including South Korea, Italy, and Iran, that are above the percolation threshold curve, and they are on the verge of COVID-19 outbreaks on Mar. 2, 2020.

The British prime minister, Boris Johnson, once suggested having a sufficiently high fraction of individuals infected by COVID-19 and recovered from the disease to achieve herd immunity. To address the question in (Q4), we argue that herd immunity corresponds to the reduction of the number of susceptible persons in the SIR model, and herd immunity can be achieved after at least $1-\frac{1}{R_0}$ fraction of individuals being infected and recovered from the COVID-19.

For (Q5), we consider two commonly used approaches for social distancing: (i) allowing every person to keep its interpersonal contacts up to a fraction of its normal contacts, and (ii) canceling mass gatherings. For the analysis of social distancing, we have to take the social network (and its network structure) into account. For this, we consider the independent cascade (IC) model for disease propagation in a random network specified by a degree distribution $p_k, k=0, 1, 2, \ldots$. The IC model has been widely used for the study of the influence maximization problem in viral marketing (see, e.g., \cite{kempe2003maximizing}). In the IC model, an {\em infected} node can transmit the disease to a neighboring {\em susceptible} node (through an edge) with a certain propagation probability. Repeatedly continuing the propagation, we have a subgraph that contains the set of infected nodes in the long run. By relating the propagation probabilities in the IC model to the transmission rates and recovering rates in the SIR model, we show two results for social distancing: (i) for the social distancing approach that allows every person to keep its interpersonal contacts up to (on average) a fraction $a$ of its normal contacts, the basic reproduction number is reduced by a factor of $a^2$, and (ii) for the social distancing approach that cancels mass gatherings by removing nodes with the number of edges larger than or equal to $k_0$, the basic reproduction number is reduced by a factor of $\displaystyle\frac{\sum_{k=0}^{k_0-2}k q_k}{\sum_{k=0}^{\infty}k q_k}$, where $q_k$ is the excess degree distribution of $p_k$.

For (Q6), there is a piece of solid evidence for an outbreak in the State of New York when Andrew Cuomo, the governor of New York State, said on Apr. 23, 2020, that $13.9\%$ of a group of $3,000$ people tested positive for COVID-19 antibodies. In the SIR model with a stationary transmission rate and a stationary recovering rate, it is well-known (see, e.g., \cite{newman2010networks}) that the ratio of the population infected in the long run, denoted by $r$, can be computed from the fixed point equation:
\beq{Poisson1234c}
1-r=e^{-R_0 r}.
\eeq
However, the SIR model does not take the network structure into account. To see the effect of the degree distribution to the ratio of the population infected in the long run, we consider the IC model for disease propagation in a random network generated by the configuration model with the degree distribution $p_k$, $k=0, 1, 2, \ldots$. We show that if $R_0>1$, then a certain proportion of the population will be infected. Moreover, the ratio of the population infected in the long run can also be computed from a fixed point equation. When the degree distribution is a Poisson degree distribution, it reduces to \req{Poisson1234c}. Our numerical results show that \req{Poisson1234c} is a conservative estimate of the ratio of the population infected in the long run in comparison with real networks that have power-law degree distributions.

In Table \ref{table1}, we provide a list of notations that are used in this paper.
\begin{table}[!htbp]
	\caption{List of notations}
	\begin{center}
		\begin{tabular}{|c|l|}
			\hline
			\textbf{Notation}&\multicolumn{1}{c|}{\textbf{Description}}\\ \hline
			$A$& The $2 \times 2$ transition matrix in the SIR model \\
			$a$ & the fraction of reduced normal contacts in social distancing \\
			$a_j$ & The $j^{th}$ coefficient of the first FIR filter\\
			$\alpha_1$ & The first regulation parameter in the ridge regression \\
			$\alpha_2$ & The second regulation parameter in the ridge regression \\
			$b_k$ & The $k^{th}$ coefficient of the second FIR filter\\
			$\beta$ & The (stationary) transmission rate \\
			$\beta(t)$ & The transmission rate at time $t$ \\
			${\hat \beta}(t)$ & The estimated/predicted transmission rate at time $t$ \\
			$\beta_1$ & The transmission rate of type \RomanNumeralCaps{1} infected persons \\
			$\beta_2$ & The transmission rate of type \RomanNumeralCaps{2} infected persons \\
			$C$ & The normalization constant of a power-law degree \\
			& \quad distribution\\
			$c$ & The average degree \\
			$g_0(z)$ & The moment generating function of the degree\\
			& distribution $p_k$\\
			$g_1(z)$ & The moment generating function of the excess \\
			&\quad degree distribution $q_k$\\
			$\gamma$ & The (stationary) recovering rate \\
			$\gamma(t)$ & The recovering rate at time $t$ \\
			${\hat \gamma}(t)$ & The estimated/predicted recovering rate at time $t$ \\
			$\gamma_1$ & The recovering rate of type \RomanNumeralCaps{1} infected persons \\
			$\gamma_2$ & The recovering rate of type \RomanNumeralCaps{2} infected persons \\
			$h$ & The probability that a randomly selected person is \\
			&\quad susceptible \\
			$n$ & The total population \\
			$\phi$ & The average propagation probability \\
			$\phi_1$ & The propagation probability of type \RomanNumeralCaps{1} infected persons \\
			$\phi_2$ & The propagation probability of type \RomanNumeralCaps{2} infected persons \\
			$p_k$ & The degree distribution \\
			$q_k$ & The excess degree distribution \\
			$R_0$ & The basic reproduction number \\
			$R_0(t)$ & The basic reproduction number at time $t$ \\
			${\hat R}_0(t)$ & The estimated/predicted basic reproduction number\\
			& \quad at time $t$ \\
			$R(t)$ & The number of recovered persons at time $t$\\
			${\hat R}(t)$ & The estimated/predicted number of recovered persons \\
			& \quad at time $t$\\
			$r$ & The ratio of the population infected in the long run\\
			$S(t)$ & The number of susceptible persons at time $t$\\
			$s$ & The reduction factor due to social distancing \\
			$T$ & The period of a historical dataset \\
			$u_1$ & The probability that the size of the infected tree of a type \RomanNumeralCaps{1}\\
			& \quad node is finite via a specific one of its neighbors\\
			$u_2$ & The probability that the size of the infected tree of a type \RomanNumeralCaps{2}\\
			& \quad node is finite via a specific one of its neighbors\\
			$v$ & The infected probability of one end node of a randomly\\
			&\quad selected edge \\
			$W$ & The prediction window \\
			$w_1$ & The probability that an infected person is of type \RomanNumeralCaps{1}\\
			$w_2$ & The probability that an infected person is of type \RomanNumeralCaps{2}\\
			$X(t)$ & The number of infected persons at time $t$\\
			${\hat X}(t)$ & The estimated/predicted number of infected persons \\
			& \quad at time $t$\\
			\hline
		\end{tabular}
		\label{table1}
	\end{center}
\end{table}

The rest of the paper is organized as follows: In Section \ref{td_sm}, we propose the time-dependent SIR model. We then extend the model to the SIR model with undetectable infected persons in Section \ref{uip_sm}. In Section \ref{IC}, we consider the independent cascade model for disease propagation in a random network specified by a degree distribution. In Section \ref{exp_r}, we conduct several numerical experiments to illustrate the effectiveness of our models. In Section \ref{discussion}, we put forward some discussions and suggestions to control COVID-19. The paper is concluded in Section \ref{conclude}.

\section{The Time-dependent SIR Model}\label{td_sm}
\subsection{Susceptible-infected-recovered (SIR) Model}
In the typical mathematical model of infectious disease, one often simplify the virus-host interaction and the evolution of an epidemic into a few basic disease states. One of the simplest epidemic model, known as the susceptible-infected-recovered (SIR) model \cite{newman2010networks}, includes three states: the susceptible state, the infected state, and the recovered state. An individual in the {\em susceptible state} is one who does not have the disease at time $t$ yet, but may be infected if one is in contact with a person infected with the disease. The {\em infected state} refers to an individual who has a disease at time $t$ and may infect a susceptible individual potentially (if they come into contact with each other). The {\em recovered state} refers to an individual who is either recovered or dead from the disease and is no longer contagious at time $t$. Also, a recovered individual will not be back to the susceptible state anymore. The reason for the number of deaths is counted in the recovered state is that, from an epidemiological point of view, this is basically the same thing, regardless of whether recovery or death does not have much impact on the spread of the disease. As such, they can be effectively eliminated from the potential host of the disease \cite{ng2001stable}. Denote by $S(t), X(t)$ and $R(t)$ the numbers of susceptible persons, infected persons, and recovered persons at time $t$. Summing up the above SIR model, we believe it is very similar to the COVID-19 outbreak, and we will adopt the SIR model as our basic model in this paper.

In the traditional SIR model, it has two time-invariant variables: the transmission rate $\beta$ and the recovering rate $\gamma$. The transmission rate $\beta$ means that each individual has on average $\beta$ contacts with randomly chosen others per unit time. On the other hand, the recovering rate $\gamma$ indicates that individuals in the infected state get recovered or die at a fixed average rate $\gamma$. The traditional SIR model neglects the time-varying property of $\beta$ and $\gamma$. This assumption is too simple to precisely and effectively predict the trend of the disease. Therefore, we propose the time-dependent SIR model, where both the transmission rate $\beta$ and the recovering rate $\gamma$ are functions of time $t$. Such a time-dependent SIR model is much better to track the disease spread, control, and predict the future trend.

\subsection{Differential Equations for the Time-dependent SIR Model}\label{CTSIR}
For the traditional SIR model, the three variables $S(t), X(t)$ and $R(t)$ are governed by the following differential equations (see, e.g., the book \cite{newman2010networks}):
\bieeeeq{traditional_SIR}
\frac{dS(t)}{dt} &=& \frac{-\beta S(t)X(t)}{n},\nonumber\\
\frac{dX(t)}{dt} &=& \frac{\beta S(t)X(t)}{n} -\gamma X(t),\nonumber\\
\frac{dR(t)}{dt} &=& \gamma X(t).\nonumber
\eieeeeq
We note that
\beq{ness_n}
S(t) + X(t) + R(t) = n,
\eeq
where $n$ is the total population. Let $\beta(t)$ and $\gamma(t)$ be transmission rate and recovering rate at time $t$. Replacing $\beta$ and $\gamma$ by $\beta(t)$ and $\gamma(t)$ in the differential equations above yields
\bieeeeq{diff_time_sir}
\frac{dS(t)}{dt} &=& \frac{-\beta (t) S(t)X(t)}{n},\label{diff_time_sir1}\\
\frac{dX(t)}{dt} &=& \frac{\beta (t) S(t)X(t)}{n} -\gamma (t) X(t),\label{diff_time_sir2}\\
\frac{dR(t)}{dt} &=& \gamma (t) X(t).\label{diff_time_sir3}
\eieeeeq
The three variables $S(t), X(t)$ and $R(t)$ still satisfy \req{ness_n}.

Now we briefly explain the intuition of these three equations. Equation (\ref{diff_time_sir1}) describes the difference of the number of susceptible persons $S(t)$ at time $t$. If we assume the total population is $n$, then the probability that a randomly chosen person is in the susceptible state is $S(t)/n$. Hence, an individual in the infected state will contact (on average) $\beta(t)S(t)/n$ people in the susceptible state per unit time, which implies the number of newly infected persons is $\beta(t)S(t)X(t)/n$ (as there are $X(t)$ people in the infected state at time $t$). On the contrary, the number of people in the susceptible state will decrease by $\beta(t)S(t)X(t)/n$. Additionally, as every individual in the infected state will recover with rate $\gamma (t)$, there are (on average) $\gamma(t)X(t)$ people recovered at time $t$. This is shown in (\ref{diff_time_sir3}) that illustrates the difference of $R(t)$ at time $t$. Since three variables $S(t), X(t)$ and $R(t)$ still satisfy \req{ness_n}, we have $$\frac{dX(t)}{dt}=-(\frac{dS(t)}{dt}+\frac{dR(t)}{dt}),$$ which is the number of people changing from the susceptible state to the infected state minus the number of people changing from the infected state to the recovered state (see (\ref{diff_time_sir2})).

\subsection{Discrete Time Time-dependent SIR Model}\label{DTSIR}
Due to the COVID-19 data is updated in days \cite{outbreak_notification_2020}, we revise the differential equations in (\ref{diff_time_sir1}), (\ref{diff_time_sir2}), and (\ref{diff_time_sir3}) into discrete time difference equations:
\bieeeeq{D_diff_time_sir_1}
S(t+1) - S(t) &=& \frac{-\beta (t)S(t)X(t)}{n},\label{ds}\\
X(t+1) - X(t) &=& \frac{\beta (t)S(t)X(t)}{n} - \gamma(t) X(t),\label{dx_pre}\\
R(t+1) - R(t) &=& \gamma (t) X(t).\label{dr}
\eieeeeq
Again, the three variables $S(t), X(t)$ and $R(t)$ still satisfy \req{ness_n}.

In the beginning of the disease spread, the number of confirmed cases is very low, and most of the population are in the susceptible state. Hence, for our analysis of the initial stage of COVID-19, we assume $\{S(t)\approx n,\ t\ge 0\}$, and further simplify (\ref{dx_pre}) as follows:
\begin{equation}\label{dx_pre2}
	X(t+1) - X(t) = \beta(t)X(t) - \gamma(t)X(t).
\end{equation}
From the difference equations above, one can easily derive $\beta(t)$ and $\gamma(t)$ of each day. From (\ref{dr}), we have
\beq{d_gamma}
\gamma (t) = \frac{R(t+1) - R(t)}{X(t)}.
\eeq
Using (\ref{dr}) in (\ref{dx_pre2}) yields
\beq{d_beta}
\beta (t) = \frac{[X(t+1)-X(t)]+[R(t+1) - R(t)]}{X(t)}.
\eeq

Given the historical data from a certain period $\{X(t),R(t),\ 0\le t \le T-1\}$, we can measure the corresponding $\{\beta (t),\gamma (t),0\le t \le T-2\}$ by using \req {d_gamma} and \req{d_beta}. With the above information, we can use machine learning methods to predict the time varying transmission rates and recovering rates.

\subsection{Tracking Transmission Rate $\beta(t)$ and Recovering Rate $\gamma(t)$ by Ridge Regression}\label{fir_ridge}
In this subsection, we track and predict $\beta(t)$ and $\gamma(t)$ by the commonly used Finite Impulse Response (FIR) filters in linear systems. Denote by $\hat{\beta} (t)$ and $\hat{\gamma} (t)$ the {\em predicted} transmission rate and recovering rate. From the FIR filters, they are predicted as follows:
\bieeeeq{fir1}
\hat{\beta} (t) &=& a_1 \beta (t-1) + a_2 \beta (t-2) + \cdots + a_J \beta (t-J) + a_0\nonumber\\
&=&\sum_{j=1}^{J}a_j \beta(t-j)+a_0,\label{beta_est}\\
\hat{\gamma} (t) &=& b_1 \gamma (t-1) + b_2 \gamma (t-2) + \cdots + b_K \gamma (t-K) + b_0\nonumber\\
&=&\sum_{k=1}^{K}b_k \gamma(t-k)+b_0,\label{gamma_est}
\eieeeeq
where $J$ and $K$ are the orders of the two FIR filters ($0< J, K < T-2$), $a_j, j=0,1,\dots,J$, and $b_k, k=0,1,\dots,K$ are the coefficients of the impulse responses of these two FIR filters.

There are several widely used machine learning methods for the estimation of the coefficients of the impulse response of an FIR filter, e.g., ordinary least squares (OLS), regularized least squares (i.e., ridge regression), and partial least squares (PLS) \cite{dayal1996identification}. In this paper, we choose the ridge regression as our estimation method that solves the following optimization problem:
\bieeeeq{ridge}
\min\limits_{a_j}\sum_{t=J}^{T-2}(\beta(t) - \hat{\beta} (t))^2 + \alpha_{1} \sum_{j=0}^{J}a_j^2,\label{ridge1}\\
\min\limits_{b_k}\sum_{t=K}^{T-2}(\gamma(t) - \hat{\gamma} (t))^2 + \alpha_{2} \sum_{k=0}^{K}b_k^2,\label{ridge2}
\eieeeeq
where $\alpha_{1}$ and $\alpha_{2}$ are the regularization parameters.

\subsection{Tracking the Number of Infected Persons $\hat{X}(t)$ and the Number of Recovered Persons $\hat{R}(t)$ of the Time-dependent SIR Model}\label{trackSIR}
In this subsection, we show how we use the two FIR filters to track and predict the number of infected persons and the number of recovered persons in the time-dependent SIR model. Given a period of historical data $\{X(t),R(t),\ 0\le t \le T-1\}$, we first measure $\{\beta (t),\gamma (t),\ 0\le t \le T-2\}$ by \req {d_gamma} and \req{d_beta}. Then we solve the ridge regression (with the objective functions in (\ref{ridge1}) and (\ref{ridge2}) and the constraints in (\ref{beta_est}) and (\ref{gamma_est})) to learn the coefficients of the FIR filters, i.e., $a_j, j=0,1,\dots,J$ and $b_k, k=0,1,\dots,K$. Once we learn these coefficients, we can predict $\hat{\beta}(t)$ and $\hat{\gamma}(t)$ at time $t=T-1$ by the trained ridge regression in (\ref{beta_est}) and (\ref{gamma_est}).

Denote by $\hat{X}(t)$ (resp. $\hat{R}(t)$) the predicted number of infected (resp. recovered) persons at time $t$. To predict $\hat{X}(t)$ and $\hat{R}(t)$ at time $t=T$, we simply replace $\beta(t)$ and $\gamma(t)$ by ${\hat \beta}(t)$ and ${\hat \gamma}(t)$ in (\ref{dr}) and (\ref{dx_pre2}). This leads to
\bieeeeq{xrT_est}
\hat{X}(T) &=& \big(1+\hat{\beta}(T-1)-\hat{\gamma}(T-1)\big)X(T-1),\label{xst_est}\\
\hat{R}(T) &=& R(T-1) +\hat{\gamma}(T-1)X(T-1).\label{rst_est}
\eieeeeq

To predict $\hat{X}(t)$ and $\hat{R}(t)$ for $t >T$, we estimate ${\hat \beta}(t)$ and $\hat{\gamma}(t)$ by using (\ref{beta_est}) and (\ref{gamma_est}). Similar to those in (\ref{xst_est}) and (\ref{rst_est}), we predict $\hat{X}(t)$ and $\hat{R}(t)$ as follows:
\bieeeeq{xr_est}
\hat{X}(t+1) &=& \big(1+\hat{\beta}(t)-\hat{\gamma}(t)\big)\hat{X}(t), \ t \ge T,\label{x_est}\\
\hat{R}(t+1) &=& \hat{R}(t) +\hat{\gamma}(t)\hat{X}(t), \ t \ge T.\label{r_est}
\eieeeeq

The detailed steps of our tracking/predicting method are outlined in Algorithm \ref{SIR_model_algo}.
\begin{algorithm}[!htbp]
	\caption{Tracking Discrete Time Time-dependent SIR Model}
	\KwIn {$\{X(t), R(t),\ 0\le t \le T-1\}$, Regularization parameters $\alpha_1$ and $\alpha_2$, Order of FIR filters $J$ and $K$, Prediction window $W$.}
	\KwOut {$\{\beta (t),\gamma (t),\ 0\le t \le T-2\}$, $\{\hat{\beta} (t),\hat{\gamma} (t),\ t \ge T-1\}$, and $\{\hat{X}(t),\hat{R}(t),\ t \ge T\}$.}
	\begin{algorithmic}[1]\label{SIR_model_algo}
		\STATE {Measure $\{\beta (t),\gamma (t),\ 0\le t \le T-2\}$ using \req{d_beta} and \req{d_gamma} respectively.}
		\STATE {Train the ridge regression using (\ref{ridge1}) and (\ref{ridge2}).}
		\STATE {Estimate $\hat{\beta}(T-1)$ and $\hat{\gamma}(T-1)$ by (\ref{beta_est}) and (\ref{gamma_est}) respectively.}
		\STATE {Estimate the number of infected persons $\hat{X}(T)$ and recovered persons $\hat{R}(T)$ on the next day $T$ using (\ref{xst_est}) and (\ref{rst_est}) respectively.}
		\WHILE {$T \le t \le T + W$}
		\STATE {Estimate $\hat{\beta}(t)$ and $\hat{\gamma}(t)$ in (\ref{beta_est}) and (\ref{gamma_est}) respectively.}
		\STATE {Predict $\hat{X}(t+1)$ and $\hat{R}(t+1)$ using (\ref{x_est}) and (\ref{r_est}) respectively.}
		\ENDWHILE
	\end{algorithmic}
\end{algorithm}

We note that this {\em deterministic} epidemic model is based on the mean-field approximation for $X(t)$ and $R(t)$. Such an approximation is a result of the law of large numbers. Therefore, when $X(t)$ and $R(t)$ are relatively small, the mean-field approximation may not be as accurate as expected. In those cases, one might have to resort to stochastic epidemic models, such as Markov chains.

\section{The SIR Model with Undetectable Infected Persons}\label{uip_sm}
According to the recent report from WHO \cite{who}, only $87.9\%$ of COVID-19 patients have a fever, and $67.7\%$ of them have a dry cough. This means there exist asymptomatic infections. Recent studies in \cite{maier2020effective} and \cite{ganyani2020estimating} also pointed out the existence of the asymptomatic carriers of COVID-19. Those people are unaware of their contagious ability, and thus get more people infected. The transmission rate can increase dramatically in this circumstance.

To take the undetectable infected persons into account, we propose the SIR model with undetectable infected persons in this section. We assume that there are two types of infected persons. The individuals who are detectable (with obvious symptoms) are categorized as type \RomanNumeralCaps{1} infected persons, and the asymptomatic individuals who are undetectable are categorized as type \RomanNumeralCaps{2} infected persons. For an infected individual, it has probability $w_1$ to be type \RomanNumeralCaps{1} and probability $w_2$ to be type \RomanNumeralCaps{2}, where $w_1+w_2=1$. Besides, those two types of infected persons have different transmission rates and recovering rates, depending on whether they are under treatment or isolation or not. We denote $\beta_1(t)$ and $\gamma_1(t)$ as the transmission rate and the recovering rate of type \RomanNumeralCaps{1} at time $t$. Similarly, $\beta_2(t)$ and $\gamma_2(t)$ are the transmission rate and the recovering rate for type \RomanNumeralCaps{2} at time $t$.

\subsection{The Governing Equations for the SIR Model with Undetectable Infected Persons}
Now we derive the governing equations for the SIR model with two types of infected persons. Let $X_1(t)$ (resp. $X_2(t)$) be the number of type \RomanNumeralCaps{1} (resp. type \RomanNumeralCaps{2}) infected persons at time $t$. Similar to the derivation of (\ref{dx_pre}), (\ref{dr}) in Subsection \ref{DTSIR}, we assume that $\{S(t)\approx n,\ t\ge 0\}$ in the initial stage of the epidemic and split $X(t)$ into two types of infected persons. We have the following difference equations:
\bieeeeq{diff_sir_undetect}
X_1(t+1) - X_1(t) &=& \beta_1 X_1(t) w_1 + \beta_2 X_2(t) w_1\nonumber\\&\ &-\gamma_1 X_1(t),\label{diff_sir_undetect1}\\
X_2(t+1) - X_2(t) &=& \beta_1 X_1(t) w_2 + \beta_2 X_2(t) w_2\nonumber\\ &\ &-\gamma_2 X_2(t),\label{diff_sir_undetect2}\\
R(t+1) - R(t) &=& \gamma_1 X_1(t) + \gamma_2 X_2(t),\label{diff_sir_undetect3}
\eieeeeq
where $\beta_1$, $\beta_2$, $\gamma_1$, and $\gamma_2$ are constants. It is noteworthy that those constants can also be time-dependent as we have in Section \ref{td_sm}. However, in this section, we set them as constants to focus on the effect of undetectable infected persons. Rewriting (\ref{diff_sir_undetect1}) and (\ref{diff_sir_undetect2}) in the matrix form yields the following matrix equation:
$$\begin{bmatrix}
{X_1}(t+1)\\
{X_2}(t+1)
\end{bmatrix}=
\begin{bmatrix}
1+\beta_1 w_1-\gamma_1 & \beta_2 w_1\\
\beta_1 w_2 & 1+\beta_2 w_2 - \gamma_2
\end{bmatrix}
\begin{bmatrix}
X_1(t)\\
X_2(t)
\end{bmatrix}
,$$
where $w_2=1-w_1$. Let $\bf A$ be the transition matrix of the above system equations, i.e.,
\beq{transition_matrix}
{\bf A}=
\begin{bmatrix}
	1+\beta_1 w_1-\gamma_1 & \beta_2 w_1\\
	\beta_1 w_2 & 1+\beta_2 w_2 - \gamma_2
\end{bmatrix}.
\eeq

It is well-known (from linear algebra) such a system is stable if the spectral radius (the largest absolute value of the eigenvalue) of $\bf A$ is less than $1$. In other words, $X_1(t+1)$ and $X_2(t+1)$ will converge gradually to finite constants when $t$ goes to infinity. In that case, there will not be an outbreak. On the contrary, if the spectral radius is greater than $1$, there will be an outbreak, and the number of infected persons will grow exponentially with respect to time $t$ (at the rate of the spectral radius).

\subsection{The Basic Reproduction Number}\label{basic}
To further examine the stability condition of such a system, we let
\beq{stab1111}
R_0=w_1 \frac{\beta_1}{\gamma_1}+w_2 \frac{\beta_2}{\gamma_2}.
\eeq
Note that $R_0$ is simply the basic reproduction number of a newly infected person as an infected person can further infect on average $\beta_1/\gamma_1$ (resp. $\beta_2/\gamma_2$) persons if it is of type \RomanNumeralCaps{1} (resp. type \RomanNumeralCaps{2}) and that happens with probability $w_1$ (resp. $w_2$). In the following theorem, we show that there is no outbreak if $R_0<1$ and there is an outbreak if $R_0>1$. Thus, $R_0$ in \req{stab1111} is known as the percolation threshold for an outbreak in such a model \cite{newman2010networks}.

\bthe{stability}
If $R_0 <1$, then the spectral radius of $\bf A$ in \req{transition_matrix} is less than $1$ and there is no outbreak of the epidemic. On the other hand, if $R_0 >1$, then the spectral radius of $\bf A$ in \req{transition_matrix} is larger than $1$ and there is an outbreak of the epidemic.
\ethe

\bproof\ (\rthe{stability})

First, we note that $\gamma_1$ and $\gamma_2$ are recovering rates and they cannot be larger than $1$ in the discrete-time setting, i.e., it takes at least one day for an infected person to recover. Thus, the matrix ${\bf A}$ is a positive matrix (with all its elements being positive). It then follows from the Perron-Frobenius theorem that the spectral radius of the matrix is the larger eigenvalue of the $2\times 2$ matrix.

Now we find the larger eigenvalue of the matrix ${\bf A}$. Let ${\bf I}$ be the $2 \times 2$ identify matrix and
\beq{stab0011}
{\tilde {\bf A}}={\bf A}-{\bf I}.
\eeq
Then
\beq{transition_matrix2}
{\tilde {\bf A}}=
\begin{bmatrix}
	\beta_1 w_1-\gamma_1 & \beta_2 w_1\\
	\beta_1 w_2 & \beta_2 w_2 - \gamma_2
\end{bmatrix}.
\eeq
Let
\beq{stab1122a}
z_1=\beta_1 w_1-\gamma_1+\beta_2 w_2 -\gamma_2,
\eeq
and
\beq{stab1122b}
z_2=\beta_1 w_1 \gamma_2+\beta_2 w_2\gamma_1-\gamma_1\gamma_2.
\eeq
It is straightforward to show that the two eigenvalues of ${\tilde {\bf A}}$ are
\beq{stab2222a}
\lambda_1 =\frac{1}{2}(z_1+\sqrt{z_1^2+4z_2}),
\eeq
and
\beq{stab2222b}
\lambda_2 =\frac{1}{2}(z_1-\sqrt{z_1^2+4z_2}).
\eeq
Note that $\lambda_1 \ge \lambda_2$. In view of \req{stab0011}, the larger eigenvalue of the transition matrix ${\bf A}$ is $1+\lambda_1$.

If $R_0 <1$, we know that $z_2 <0$, $w_1 \frac{\beta_1}{\gamma_1} <1$, and $w_2 \frac{\beta_2}{\gamma_2} <1$. Thus, we have from \req{stab1122a} that $z_1<0$. In view of \req{stab2222a}, we conclude that $$\lambda_1 < \frac{1}{2} (z_1 + |z_1|) =0.$$ This shows that $1+\lambda_1<1$ and the spectral radius of $\bf A$ is less than $1$.

On the other hand, if $R_0 >1$, then $z_2 >0$ and we have from \req{stab2222a} that $$\lambda_1 > \frac{1}{2} (z_1 + |z_1|) \ge 0.$$ This shows that $1+\lambda_1>1$ and the spectral radius of $\bf A$ is larger than $1$.
\eproof

The relation between the system parameters and the phase transition will be shown in Subsection \ref{perco_experiment}.

\subsection{Herd Immunity}\label{herd}
Herd immunity is one way to resist the spread of a contagious disease if a sufficiently high fraction of individuals are immune to the disease, especially through vaccination. One interesting strategy, once suggested by Boris Johnson, the British prime minister, is to have a sufficiently high fraction of individuals infected by COVID-19 and recovered from the disease to achieve herd immunity. The question is, what will be the fraction of individuals that need to be infected to achieve herd immunity for COVID-19.

To address such a question, we note that herd immunity corresponds to the reduction of the number of susceptible persons in the SIR model. In our previous analysis, we all assume that every person is susceptible to COVID-19 at the early stage and thus $S(t)/n \approx 1$. For the analysis of herd immunity, we assume that there is a probability $h$ that a randomly chosen person is susceptible at time $t$. This is equivalent to that $1-h$ fraction of individuals are immune to the disease. Under such an assumption, we then have
\beq{herd1111}
\frac{S(t)}{n} \approx h.
\eeq

In view of the difference equation for $X(t)$ in (\ref{dx_pre}), we can rewrite (\ref{diff_sir_undetect1})-(\ref{diff_sir_undetect3}) to derive the governing equations for herd immunity as follows:
\bieeeeq{diff_sir_undetecth}
X_1(t+1) - X_1(t) &=& \beta_1 X_1(t)h w_1 + \beta_2 X_2(t)h w_1\nonumber\\&\ &-\gamma_1 X_1(t),\label{diff_sir_undetecth1}\\
X_2(t+1) - X_2(t) &=& \beta_1 X_1(t)h w_2 + \beta_2 X_2(t)h w_2\nonumber\\ &\ &-\gamma_2 X_2(t),\label{diff_sir_undetecth2}\\
R(t+1) - R(t) &=& \gamma_1 X_1(t) + \gamma_2 X_2(t).\label{diff_sir_undetecth3}
\eieeeeq
In comparison with the original governing equations in (\ref{diff_sir_undetect1})-(\ref{diff_sir_undetect3}), the only difference is the change of the transmission rate of type \RomanNumeralCaps{1} (resp. type \RomanNumeralCaps{2}) from $\beta_1$ to $\beta_1 h$ (resp. from $\beta_2$ to $\beta_2 h$). Thus, herd immunity effectively reduces the transmission rates by a factor of $h$. As a direct consequence of \rthe{stability}, we have the following corollary.

\bcor{herd}
For a contagious disease modeled by our SIR model with two types of infected persons that has $R_0$ in \req{stab1111} greater than $1$,
herd immunity can be achieved after at least $1-h^*$ fraction of individuals being infected and recovered from the contagious disease, where $$h^*=\frac{1}{R_0}.$$
\ecor

\section{The Independent Cascade (IC) Model for Disease Propagation in Networks}\label{IC}
Our analysis in the previous section does not consider how the structure of a social network affects the propagation of a disease. There are other widely used policies, such as social distancing, that could not be modeled by our SIR model with undetectable infected persons in Section \ref{uip_sm}. To take the network structure into account, in this section, we consider the independent cascade (IC) model for disease propagation. The IC model was previously studied by Kempe, Kleinberg, and Tardos in \cite{kempe2003maximizing} for the influence maximization problem in viral marketing. In the IC model, there is a social network modeled by a graph $G=(V,E)$, where $V$ is the set of nodes, and $E$ is the set of edges. An {\em infected} node can transmit the disease to a neighboring {\em susceptible} node (through an edge) with a certain propagation probability. As there are two types of infected persons in our model, we denote by $\phi_1$ (resp. $\phi_2$) the propagation probability that a type \RomanNumeralCaps{1} (resp. type \RomanNumeralCaps{2}) infected node transmits the disease to an (immediate) neighbor of the infected node. Once a neighboring node is infected, it becomes a type \RomanNumeralCaps{1} (type \RomanNumeralCaps{2}) infected node with probability $w_1$ (resp. $w_2)$ and it can continue the propagation of the disease to its neighbors. Continuing the propagation, we thus form a subgraph of $G$ that contains the set of infected nodes in the long run. Call such a subgraph the {\em infected subgraph}. One interesting question is how one controls the spread of the disease so that the total number of nodes in the infected subgraph remains small even when the total number of nodes is very large.

\subsection{The Infected Tree in the Configuration Model}
The exact network structure, i.e., the adjacency matrix of the network $G$, is in general very difficult to obtain for a large population. However, it might be possible to learn some characteristics of the network, in particular, the degree distribution of the nodes. The configuration model (see, e.g., the book \cite{newman2010networks}) is one family of random networks that are specified by degree distributions of nodes. A randomly selected node in such a random network has degree $k$ with probability $p_k$. The edges of a node are randomly connected to the edges of the other nodes. As the edge connections are random, the infected subgraph appears to be a {\em tree} (with high probability) if one follows an edge of an infected node to propagate the disease to the other nodes in such a network. The tree assumption is one of the most important properties of the configuration model. Another crucial property of the configuration model is the {\em excess degree distribution}. The probability that one finds a node with degree $k+1$ along an edge connected to that node is
\beq{excess1111}
q_k=\frac{(k+1)p_{k+1}}{\sum_{\ell=0}^\infty(\ell+1)p_{\ell+1}}.
\eeq
Thus, excluding the edge coming to the node, there are still $k$ edges that can propagate the disease. This is also the reason why $q_k$ is called the excess degree distribution. Note that the excess distribution $q_k$ is in general different from the degree distribution $p_k$. They are the same when $p_k$ is the Poisson degree distribution. In that case, the configuration model reduces to the famous Erd\"os-R\'enyi random graph.

As the infected subgraph is a tree in the configuration model, we are interested to know whether the size of the infected tree is finite. We say that there is no outbreak if the size of the infected tree of an infected node is finite with probability $1$. Let $u_1$ (resp. $u_2$) be the probability that the size of the infected tree of a type \RomanNumeralCaps{1} (resp. type \RomanNumeralCaps{2}) node is finite via a specific one of its neighbors. Then,
\bieeeeq{u_def}
u_1 = 1-\phi_1 + \phi_1 \sum\limits_{k=0}^{\infty}(w_1 q_k u_1^k + w_2 q_k u_2^k),\label{u1_ori}\\
u_2 = 1-\phi_2 + \phi_2 \sum\limits_{k=0}^{\infty}(w_1 q_k u_1^k + w_2 q_k u_2^k).\label{u2_ori}
\eieeeeq
To see the intuition of (\ref{u1_ori}), we note that either the neighbor is infected or not infected. It is not infected with probability $1-\phi_1$. On the other hand, it is infected with probability $\phi_1$. Then with probability $w_1$ (resp. $w_2$), the infected neighbor is of type \RomanNumeralCaps{1} (reps. type \RomanNumeralCaps{2}). Also, with probability $q_k$, the neighbor has additional $k$ edges to transmit the disease. From the tree assumption, the probability that these $k$ edges all have finite infected trees is $u_1^k$ (resp. $u_2^k$) if the infected neighbor is of type \RomanNumeralCaps{1} (reps. type \RomanNumeralCaps{2}). The equation in (\ref{u2_ori}) follows from a similar argument.

Let
\beq{moment1111}
g_1(z) = \sum\limits_{k=0}^{\infty} q_k z^k
\eeq
be the moment generating function of the excess degree distribution. Then we can simplify (\ref{u1_ori}) and (\ref{u2_ori}) as follows:
\bieeeeq{u_def_g}
u_1 = 1-\phi_1 + \phi_1 w_1 g_1(u_1) + \phi_1 w_2 g_1(u_2),\label{gov_u1}\\
u_2 = 1-\phi_2 + \phi_2 w_1 g_1(u_1) + \phi_2 w_2 g_1(u_2).\label{gov_u2}
\eieeeeq

From (\ref{gov_u1}) and (\ref{gov_u2}), we can solve $u_1$ and $u_2$ by starting from $u_1^{(0)} = u_2^{(0)} = 0$ and
\bieeeeq{sol_u1u2}
u_1^{(m+1)} = 1-\phi_1 + \phi_1 w_1 g_1(u_1^{(m)}) + \phi_1 w_2 g_1(u_2^{(m)}),\\
u_2^{(m+1)} = 1-\phi_2 + \phi_2 w_1 g_1(u_1^{(m)}) + \phi_2 w_2 g_1(u_2^{(m)}).
\eieeeeq

It is easy to show (by induction) that $u_1^{(m+1)} \ge u_1^{(m)}$ and $u_2^{(m+1)} \ge u_2^{(m)}$. Thus, they converge to some fixed point solution $u_1^*$ and $u_2^*$ of (\ref{gov_u1}) and (\ref{gov_u2}).

\subsection{Connections to the Previous SIR Model}
Now we show the connections to the SIR model in Section \ref{uip_sm} by specifying the propagation probabilities $\phi_1$ and $\phi_2$.

Suppose that one end of a randomly selected edge is a type \RomanNumeralCaps{1} node. Then this type \RomanNumeralCaps{1} node will infect $\beta_1 / \gamma_1$ persons on average from the SIR model in Section \ref{uip_sm}. Since the average excess degree is $\sum_{k=0}^{\infty} k q_k=g_1^\prime(1)$, the average number of neighbors infected by this type \RomanNumeralCaps{1} node is $\phi_1 g_1^\prime(1)$. In order for a type \RomanNumeralCaps{1} node to infect the same average number of nodes in the SIR model in Section \ref{uip_sm}, we have
\begin{equation}
	\phi_1 = \frac{\beta_1/\gamma_1}{g_1^\prime(1)}.\label{phi_1}
\end{equation}
Similarly for $\phi_2$, we have
\begin{equation}
	\phi_2 = \frac{\beta_2/\gamma_2}{g_1^\prime(1)}.\label{phi_2}
\end{equation}

With the propagation probabilities $\phi_1$ and $\phi_2$ specified in (\ref{phi_1}) and (\ref{phi_2}), we have the following stability result.

\bthe{stabilityIC}
For the IC model (for disease propagation) in a random network constructed by the configuration model, suppose that the propagation probabilities $\phi_1$ and $\phi_2$ are specified in (\ref{phi_1}) and (\ref{phi_2}). Then the size of the infected tree is finite with probability $1$ if $$R_0=w_1 \frac{\beta_1}{\gamma_1} + w_2 \frac{\beta_2}{\gamma_2}<1.$$ Under such a condition, there is no outbreak.
\ethe

\bproof\ (\rthe{stabilityIC})

Let ${\bf u}=(u_1,u_2)^T$ and ${\bf e}=(1,1)^T$. It suffices to show that ${\bf u}={\bf e}$ is the unique solution for the system of equations in
(\ref{gov_u1}) and (\ref{gov_u2}) if $R_0<1$. We prove this by contradiction. Suppose that there is a solution of (\ref{gov_u1}) and (\ref{gov_u2}) that either $u_1 <1$ or $u_2<1$ when $R_0<1$.

Since the moment generating function in \req{moment1111} is a convex function, we have from the first order Taylor's expansion for $g_1(u_1)$ and $g_1(u_2)$ that
\bieeeeq{TE}
g_1(u_1) &\ge& g_1(1) +(u_1-1)g_1^\prime(1),\label{T_exp_g1}\\
g_1(u_2) &\ge& g_1(1) +(u_2-1)g_1^\prime(1).\label{T_exp_g2}
\eieeeeq
Note that $g_1(1) = \sum_{k=0}^{\infty} q_k = 1$. Replacing (\ref{T_exp_g1}) and (\ref{T_exp_g2}) into (\ref{gov_u1}) and (\ref{gov_u2}), we have
\bieeeeq{u_hi_TE}
u_1 &\ge& 1+\phi_1 g_1^\prime(1)(w_1 u_1 + w_2 u_2 -1),\label{u1_hi_TE}\\
u_2 &\ge& 1+\phi_2 g_1^\prime(1)(w_1 u_1 + w_2 u_2 -1).\label{u2_hi_TE}
\eieeeeq
Writing these two equations in the matrix form yields
$$\begin{bmatrix*}
u_1\\
u_2
\end{bmatrix*}\ge
\begin{bmatrix*}
\phi_1 g_1^\prime(1) w_1 & \phi_1 g_1^\prime(1) w_2\\
\phi_2 g_1^\prime(1) w_1 & \phi_2 g_1^\prime(1) w_2
\end{bmatrix*}
\begin{bmatrix*}
u_1\\
u_2
\end{bmatrix*}+
\begin{bmatrix*}
1-\phi_1 g_1^\prime(1)\\
1-\phi_2 g_1^\prime(1)
\end{bmatrix*}
.$$
This can be further simplified by using (\ref{phi_1}) and (\ref{phi_2}). Thus, we have

\beq{matrix_sys_eq_sim}
\begin{bmatrix*}
	u_1\\
	u_2
\end{bmatrix*}\ge
\begin{bmatrix*}
	w_1 {\beta_1}/{\gamma_1} & w_2 {\beta_1}/{\gamma_1}\\
	w_1 {\beta_2}/{\gamma_2} & w_2 {\beta_2}/{\gamma_2}
\end{bmatrix*}
\begin{bmatrix*}
	u_1\\
	u_2
\end{bmatrix*}+
\begin{bmatrix*}
	1-{\beta_1}/{\gamma_1}\\
	1-{\beta_2}/{\gamma_2}
\end{bmatrix*}
.
\eeq

Let
\begin{equation}\label{B_matrix}
	{\bf B}=
	\begin{bmatrix*}
		w_1 {\beta_1}/{\gamma_1} & w_2 {\beta_1}/{\gamma_1}\\
		w_1 {\beta_2}/{\gamma_2} & w_2 {\beta_2}/{\gamma_2}
	\end{bmatrix*},
\end{equation}
and
\begin{equation}\label{Z_vector}
	{\bf z}=
	\begin{bmatrix*}
		1-{\beta_1}/{\gamma_1}\\
		1-{\beta_2}/{\gamma_2}
	\end{bmatrix*}.
\end{equation}
We now rewrite \req{matrix_sys_eq_sim} in the following matrix form:
\beq{matrix1111}
{\bf u} \ge {\bf B}{\bf u}+{\bf z}.
\eeq
It is straightforward to see that the two eigenvalues of $\bf B$ are $${\tilde \lambda}_1=0, \;\mbox{and}\; {\tilde \lambda}_2 = w_1 \frac{\beta_1}{\gamma_1} + w_2 \frac{\beta_2}{\gamma_2}=R_0.$$
Moreover, the eigenvector corresponding to the eigenvalue $R_0$ is $${\bf v}=(\frac{\beta_1}{\gamma_1}, \frac{\beta_2}{\gamma_2})^T.$$ Recursively expanding \req{matrix1111} for $m$ times yields
\bieeeeq{matrix2222}
{\bf u} &\ge& {\bf B}^{m+1}{\bf u}+({\bf I}+{\bf B}+ \ldots +{\bf B}^{m}){\bf z}\nonumber\\
&=&{\bf B}^{m+1}{\bf u}+{\bf e}+R_0^m {\bf v}.
\eieeeeq

Since $R_0 <1$, both ${\bf B}^{m+1}{\bf u}$ and $R_0^m {\bf v}$ converge to the zero vectors as $m \to\infty$. Letting $m \to \infty$ in \req{matrix2222} yields ${\bf u} \ge {\bf e}$. This contradicts to the assumption that either $u_1 <1$ or $u_2<1$.
\eproof

\bsubsec{Social Distancing}{socialdis}

Social distancing is an effective way to slow down the spread of a contagious disease. One common approach of social distancing is to allow every person to keep its interpersonal contacts up to (on average) a fraction $a$ of its normal contacts (see, e.g., \cite{Chen1999,becker2015modeling}). In our IC model, this corresponds to that every node randomly disconnects one of its edges with probability $1-a$.

As in the previous subsection, we let $u_1$ (resp. $u_2$) be the probability that the size of the infected tree of a type \RomanNumeralCaps{1} (resp. type \RomanNumeralCaps{2}) node is finite via a specific one of its neighbors. Then,
\bieeeeq{u_defsocial}
u_1 &=& 1-a^2 \phi_1 + a^2 \phi_1 \sum_{k=0}^{\infty}(w_1 q_k u_1^k + w_2 q_k u_2^k),\label{u1_ori2}\\
u_2 &=& 1-a^2 \phi_2 + a^2 \phi_2 \sum_{k=0}^{\infty}(w_1 q_k u_1^k + w_2 q_k u_2^k).\label{u2_ori2}
\eieeeeq
To see (\ref{u1_ori2}), note that a neighboring node of an infected node can be infected only if (i) the edge connecting these two nodes is not removed (with probability $a^2$), and (ii) the disease propagates through the edge (with the propagation probability $\phi_1$). This happens with probability $a^2 \phi_1$. Then with probability $w_1$ (resp. $w_2$), the infected neighbor is of type \RomanNumeralCaps{1} (reps. type \RomanNumeralCaps{2}). Also, with probability $q_k$, the neighbor has additional $k$ edges to transmit the disease. From the tree assumption, the probability that these $k$ edges all have finite infected trees is $u_1^k$ (resp. $u_2^k$) if the infected neighbor is of type \RomanNumeralCaps{1} (reps. type \RomanNumeralCaps{2}). The equation in (\ref{u2_ori2}) follows from a similar argument.

In comparison with the two equations in (\ref{u1_ori}) and (\ref{u2_ori}), we conclude that this approach of social distancing reduces the propagation probabilities $\phi_1$ and $\phi_2$ to $a^2 \phi_1$ and $a^2 \phi_2$, respectively. As a direct consequence of \rthe{stabilityIC}, we have the following corollary.

\bcor{socialdis}
Suppose that a social distancing approach allows every person to keep its interpersonal contacts up to (on average) a fraction $a$ of its normal contacts. For the IC model (for disease propagation) in a random network constructed by the configuration model, the size of the infected tree is finite with probability $1$ if
\beq{socialdis9999}
a^2 R_0<1.
\eeq
Under such a condition, there is no outbreak.
\ecor

Another commonly used approach of social distancing is canceling mass gatherings. Such an approach aims to eliminate the effect of ``superspreaders'' who have lots of interpersonal contacts. For this, we consider a disease control parameter $k_0$ and remove nodes with the number of edges larger than or equal to $k_0$ in our IC model. Analogous to the derivation of (\ref{u1_ori}) and (\ref{u2_ori}), we have
\bieeeeq{u_ss}
u_1 &=& 1-\phi_1 +\phi_1 \sum_{k=k_0-1}^{\infty}q_k\nonumber\\
&&\qquad \quad+ \phi_1 \sum_{k=0}^{k_0-2}q_k(w_1 u_1^k + w_2 u_2^k),\label{u1_ss}\\
u_2 &=& 1-\phi_2 +\phi_2 \sum_{k=k_0-1}^{\infty}q_k\nonumber\\
&&\qquad \quad+ \phi_2 \sum_{k=0}^{k_0-2}q_k(w_1 u_1^k + w_2 u_2^k).\label{u2_ss}
\eieeeeq
To see (\ref{u1_ss}), we note that a type \RomanNumeralCaps{1} infected person only infects a finite number of persons along an edge if (i) the disease does not propagate through the edge (with probability $1-\phi_1$), (ii) the disease propagates through the edge and the neighboring node is removed (with probability $\phi_1 \sum_{k=k_0-1}^{\infty}q_k$), and (iii) the disease propagates through the edge and the neighboring node only infects a finite number of persons (with probability $\phi_1 \sum_{k=0}^{k_0-2}q_k(w_1 u_1^k + w_2 u_2^k)$). The argument for (\ref{u2_ss}) is similar.

Analogous to the stability result of \rthe{stabilityIC}, we have the following stability result for a social distancing approach that cancels mass gatherings.

\bthe{socialdisb}
Consider a social distancing approach that cancels mass gatherings by removing nodes with the number of edges larger than or equal to $k_0$. For the IC model (for disease propagation) in a random network constructed by the configuration model, suppose that the propagation probabilities $\phi_1$ and $\phi_2$ are specified in (\ref{phi_1}) and (\ref{phi_2}). Then the size of the infected tree is finite with probability $1$ if
\beq{socialdis9999b}
\Big(\frac{\sum_{k=0}^{k_0-2}k q_k}{\sum_{k=0}^{\infty}k q_k} \Big ) R_0<1.
\eeq
Under such a condition, there is no outbreak.
\ethe

\bproof\ (\rthe{socialdisb})

As in the proof of \rthe{stabilityIC}, we let ${\bf u}=(u_1,u_2)^T$ and ${\bf e}=(1,1)^T$. It suffices to show that ${\bf u}={\bf e}$ is the unique solution for the system of equations in (\ref{u1_ss}) and (\ref{u2_ss}) if the inequality in \req{socialdis9999b} is satisfied. We prove this by contradiction. Suppose that there is a solution of (\ref{u1_ss}) and (\ref{u2_ss}) that either $u_1 <1$ or $u_2<1$ when the inequality in \req{socialdis9999b} is satisfied.

Since $f(u)=u^k$ is a convex function for $u \ge 0$ and $k \ge 0$, we have $u^k \ge 1+(u-1)k$. It then follows from (\ref{u1_ss}) and (\ref{u2_ss}) that
\bieeeeq{u_hi_TEb}
u_1 &\ge& 1-\phi_1 +\phi_1 \sum_{k=k_0-1}^{\infty}q_k +\phi_1 \sum_{k=0}^{k_0-2}q_k\nonumber\\
&&\qquad \quad+\phi_1 \sum_{k=0}^{k_0-2} k q_k (w_1 u_1 + w_2 u_2 -1),\label{u1_hi_TEb}\\
u_2 &\ge& 1-\phi_2 +\phi_2 \sum_{k=k_0-1}^{\infty}q_k +\phi_2 \sum_{k=0}^{k_0-2}q_k\nonumber\\
&&\qquad \quad+\phi_2 \sum_{k=0}^{k_0-2} k q_k (w_1 u_1 + w_2 u_2 -1).\label{u2_hi_TEb}
\eieeeeq
Writing these two equations in the matrix form and using (\ref{phi_1}) and (\ref{phi_2}) yields
\bieeeeq{matrix_sys_eqb}
\begin{bmatrix*}
	u_1\\
	u_2
\end{bmatrix*}
&\ge&
(\sum_{k=0}^{k_0-2}k q_k)
\begin{bmatrix*}
	\phi_1 w_1 & \phi_1 w_2\\
	\phi_2 w_1 & \phi_2 w_2
\end{bmatrix*}
\begin{bmatrix*}
	u_1\\
	u_2
\end{bmatrix*}\nonumber\\
&&+
\begin{bmatrix*}
	1-\phi_1 (\sum_{k=0}^{k_0-2}k q_k)\\
	1-\phi_2 (\sum_{k=0}^{k_0-2}k q_k)
\end{bmatrix*}\nonumber\\
&=&\displaystyle\frac{(\displaystyle\sum_{k=0}^{k_0-2}k q_k)}{g_1^\prime(1)} {\bf B}
\begin{bmatrix*}
	u_1\\
	u_2
\end{bmatrix*}+
\begin{bmatrix*}
	1-\displaystyle\frac{(\displaystyle\sum_{k=0}^{k_0-2}k q_k)}{g_1^\prime(1)} \displaystyle\frac{\beta_1}{\gamma_1}\\
	1-\displaystyle\frac{(\displaystyle\sum_{k=0}^{k_0-2}k q_k)}{g_1^\prime(1)} \displaystyle\frac{\beta_2}{\gamma_2}
\end{bmatrix*},
\eieeeeq
where $\bf B$ is the matrix in (\ref{B_matrix}). Note that $g_1^\prime(1)=\sum_{k=0}^{\infty}k q_k$ is simply the expected excess degree. Following the same argument as that in \rthe{stabilityIC}, one can easily show that $u_1 \ge 1$ and $u_2 \ge 1$ when the inequality in \req{socialdis9999b} is satisfied. This contradicts to the assumption that either $u_1 <1$ or $u_2<1$.
\eproof

Unfortunately, it is difficult to obtain an explicit expression for $k_0$ to prevent an outbreak in \req{socialdis9999b}. For this, we will resort to numerical computations in the next section.

\bsubsec{The Ratio of the Population Infected in the Long Run}{longrun}
Andrew Cuomo, the governor of New York State, said on Apr. 23, 2020, that $13.9\%$ of a group of $3,000$ people tested positive for COVID-19 antibodies. It is even higher in the City of New York. Such a test implies that a certain proportion of the population in the State of New York were infected (and recovered), and that is a piece of solid evidence for an outbreak in the State of New York. In all our previous studies, we have been focusing on how to prevent an outbreak. If the disease cannot be contained, we ask the question of what is the ratio of the population infected in the long run.

For the SIR model, the ratio of the population infected in the long run, denoted by $r$, is defined as $$r=\lim_{t \to\infty}\frac{R(t)}{n}.$$ If the SIR model has the stationary transmission rate $\beta$ and the stationary recovering rate $\gamma$, then the basic reproduction number $R_0$ is simply $\beta/\gamma$. It is well-known (see, e.g., \cite{newman2010networks}) that $r$ in the stationary SIR model satisfies the fixed point equation:
\beq{Poisson1234b}
1-r=e^{-R_0 r}.
\eeq

As discussed in the previous section, the SIR model does not take the network structure into account. To see the effect of the degree distribution to the ratio of the population infected in the long run, we consider the IC model for disease propagation in a random network generated by the configuration model with the degree distribution $p_k$, $k=0, 1, 2, \ldots$. In \rthe{stabilityIC}, we already showed that there is no outbreak if $R_0=w_1 \frac{\beta_1}{\gamma_1} + w_2 \frac{\beta_2}{\gamma_2}<1$. In this subsection, we will show that if $R_0>1$, then a certain proportion of the population will be infected. This is stated in the following theorem.

\bthe{longrun}
For the IC model (for disease propagation) in a random network constructed by the configuration model, suppose that the propagation probabilities $\phi_1$ and $\phi_2$ are specified in (\ref{phi_1}) and (\ref{phi_2}). Let $g_0(z)=\sum_{k=0}^\infty p_k z^k$ (resp. $g_1(z)=\sum_{k=0}^\infty q_k z^k$) be the moment generating function of the degree distribution (resp. excess degree distribution). If $$R_0=w_1 \frac{\beta_1}{\gamma_1} + w_2 \frac{\beta_2}{\gamma_2}>1,$$ then there is a nonzero probability $r$ that a randomly selected node is infected in the long run (after the propagation of the disease in the IC model). The probability $r$ can be computed by the following equation:
\beq{long3300}
r= 1-g_0(1-v +v(1-\phi)),
\eeq
where
\beq{long1122}
\phi=w_1\phi_1 +w_2 \phi_2.
\eeq
and $v$ is the probability that one end node of a randomly selected edge is infected by one of its neighbors that is not on the other end of the selected edge. Moreover, the probability $v$ is the unique solution in $(0,1]$ of the following fixed point equation:
\beq{long2222}
1-v= g_1(1-v +v(1-\phi)).
\eeq
\ethe

\bproof\ (\rthe{longrun})

\begin{figure}[!htbp]
	\centering
	\includegraphics[width=0.25\textwidth]{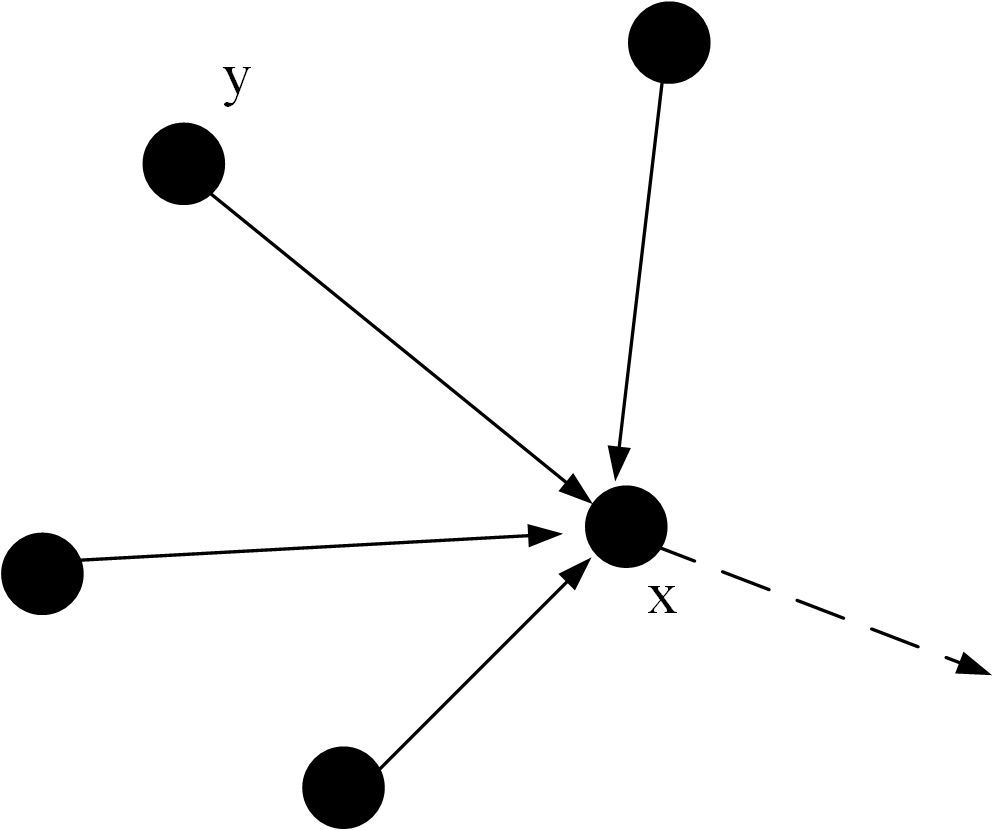}
	\caption{An illustration of the derivation of \req{long1111}. Node $x$ is on one end of a randomly selected edge, and node $y$ is an (excess) neighbor of $x$. Here the number of excess neighbors $k$ is 4.}
	\label{fig:excess}
\end{figure}

We first show \req{long2222}. Using \req{long1122}, we can rewrite \req{long2222} as follows:
\beq{long1111}
1-v= \sum_{k=0}^\infty q_k \Big (1-v +v(w_1(1-\phi_1)+w_2(1-\phi_2)) \Big )^k.
\eeq
To explain \req{long1111}, let us consider a node, say node $x$, that is on one end of a randomly selected edge (see \rfig{excess} for an illustration). Excluding the neighbor on the other end of the randomly selected edge, we call the remaining neighbors of $x$ its {\em excess neighbors} (like the excess degree distribution). The probability $1-v$ on the left-hand side of \req{long1111} is simply the probability that node $x$ is not infected by one of its excess neighbors. Suppose that there are $k$ excess neighbors of $x$. An excess neighbor of $x$, say node $y$, is on the other end of an edge and it is not infected with probability $v$. If node $y$ is not infected, then $y$ cannot infect $x$. On the other hand, if $y$ is infected, then with probability $w_1$ (resp. $w_2$) $y$ is of type \RomanNumeralCaps{1} (resp. \RomanNumeralCaps{2}) and it does not infect $x$ with the probability $1-\phi_1$ (resp. $1-\phi_2$). Thus, the probability that node $x$ is not infected from node $y$ is $1-v +v(w_1(1-\phi_1)+w_2(1-\phi_2))$. As there are $k$ (excess) neighbors, node $x$ is not infected is $\Big (1-v +v(w_1(1-\phi_1)+w_2(1-\phi_2)) \Big )^k$ from the tree assumption (in the configuration model). Since the probability that there are $k$ excess neighbors of node $x$ is $q_k$, averaging over the excess degree distribution yields \req{long1111}.

Similarly, we have
\bieeeeq{long3333}
1-r&=& \sum_{k=0}^\infty p_k \Big (1-v +v(w_1(1-\phi_1)+w_2(1-\phi_2)) \Big )^k\nonumber\\
&=& g_0(1-v +v(1-\phi)).
\eieeeeq
This then leads to \req{long3300}.

With $\phi_1$ and $\phi_2$ being specified in (\ref{phi_1}) and (\ref{phi_2}) and $R_0$ in \req{imp1111}, we have
\beq{long4444}
\phi=\frac{R_0}{g_1^\prime(1)}.
\eeq

To find $r$, we need to solve $v$ from the fixed point equation in \req{long2222}. For this, we let ${\tilde v}=1-v$ and rewrite \req{long2222} as follows:
\beq{long2222b}
{\tilde v}= g_1({\tilde v} +(1-{\tilde v})(1-\phi)).
\eeq
Since $g_1(1)=1$ and $g_1(u)$ is a convex function, $$g_1(u) \ge g_1(1)+(u-1) g_1^\prime (1)=1+(u-1)g_1^\prime (1).$$ Analogous to the argument in the proofs of \rthe{stabilityIC} and \rthe{socialdisb}, it is easy to see from \req{long4444} and \req{long2222b} that ${\tilde v}=1$ (and thus $v=0$) if $R_0 \le 1$. Moreover, there is a unique ${\tilde v}<1$ (and thus $v>0$) for the fixed point equation in \req{long2222b} if $R_0 >1$. Such a solution can be solved iteratively by using the initial value 0 for ${\tilde v}$ as described in \req{sol_u1u2}.
\eproof

In particular, for the ER model, the degree distribution $p_k$ is the Poisson degree distribution with mean $c$, i.e., $$p_k=\frac{e^{-c} c^k}{k!}.$$ The excess degree distribution of the Poisson degree distribution is the same as the degree distribution, i.e., $q_k=p_k$. In this case, we have $r=v$ and $r$ is the unique solution in $(0,1]$ of the fixed point equation
\beq{Poisson1234}
1-r=e^{-R_0 r},
\eeq
when $R_0 >1$. This is exactly the same as \req{Poisson1234b} for the SIR model with the stationary transmission rate $\beta$ and the stationary recovering rate $\gamma$. The fixed point equation in \req{Poisson1234} has a very intuitive explanation. The left-hand side is the probability that a randomly selected node is not infected, while the right-hand side is the probability that an infected node does not infect any of its neighbors as the degree distribution is Poisson.

\section{Numerical Results}\label{exp_r}
\subsection{Dataset}
In this section, we analyze and predict the trend of COVID-19 by using our time-dependent SIR model in Section \ref{td_sm} and the SIR model with undetectable infected persons in Section \ref{uip_sm}. For our analysis and prediction of COVID-19, we collect our dataset from the National Health Commission of the People’s Republic of China (NHC) daily Outbreak Notification \cite{outbreak_notification_2020}. NHC announces the data as of 24:00 the day before. We collect the number of confirmed cases, the number of recovered persons, and the number of deaths from Jan. 15, 2020 to Mar. 2, 2020 as our dataset. The confirmed case is defined as the individual with positive real-time reverse transcription polymerase chain reaction (rRT-PCR) result. It is worth noting that in the Hubei province, the definition of the confirmed case has been relaxed to the clinical features since Feb. 12, 2020, while the other provinces use the same definition as before.

\subsection{Parameter Setup}
For our time-dependent SIR model, we set the orders of the FIR filters for predicting $\beta(t)$ and $\gamma(t)$ as $3$, i.e., $J=K=3$. The stopping criteria of the model is set to $X(t)\le 0$. Since the numbers of infected persons before Jan. 27, 2020 are too small to exhibit a clear trend (which may contain noises), we only use the data after Jan. 27, 2020 as our training data for predicting $\beta(t)$ and $\gamma(t)$.

We use the scikit-learn library \cite{scikit-learn} (a third-party library of Python 3) to compute the ridge regression. The regularization parameters of predicting $\beta(t)$ and $\gamma(t)$ are set to $0.03$ and $10^{-6}$ respectively. Since the transmission rate $\beta(t)$ is nonnegative, we set it to $0$ if it is less than $0$. Then, we use Algorithm \ref{SIR_model_algo} to predict the trend of COVID-19.

\subsection{Time Evolution of the Time-dependent SIR Model}
\begin{figure}[!htbp]
	\centering
	\includegraphics[width=0.5\textwidth]{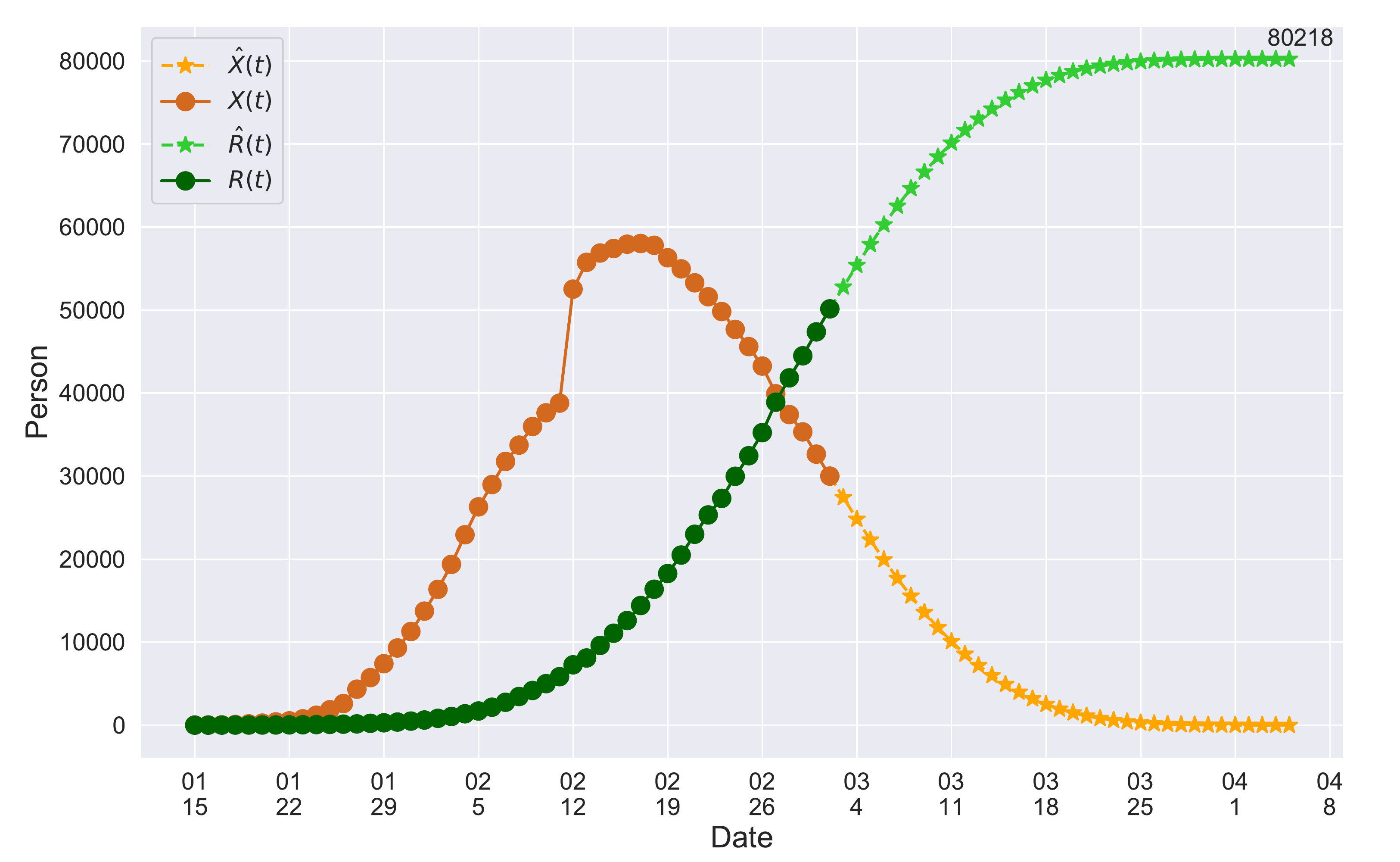}
	\caption{Time evolution of the time-dependent SIR model of the COVID-19. The circle-marked solid curve with dark orange (resp. green) color is the real number of infected persons $X(t)$ (resp. recovered persons $R(t)$), the star-marked dashed curve with light orange (resp. green) color is the predicted number of infected persons $\hat{X}(t)$ (resp. recovered $\hat{R}(t)$ persons).}
	\label{SIR_f1}
\end{figure}

In Figure \ref{SIR_f1}, we show the time evolution of the number of infected persons and the number of recovered persons. The {\em circle-marked solid curves} are the real historical data by Mar. 2, 2020, and the {\em star-marked dashed curves} are our prediction results for the future. The prediction results imply that the disease will end in $6$ weeks, and the number of the total confirmed cases would be roughly $80,000$ if the Chinese government remains their control policy, such as city-wide lockdown and suspension of works and classes.

\begin{figure}[!htbp]
	\centering
	\includegraphics[width=0.48\textwidth]{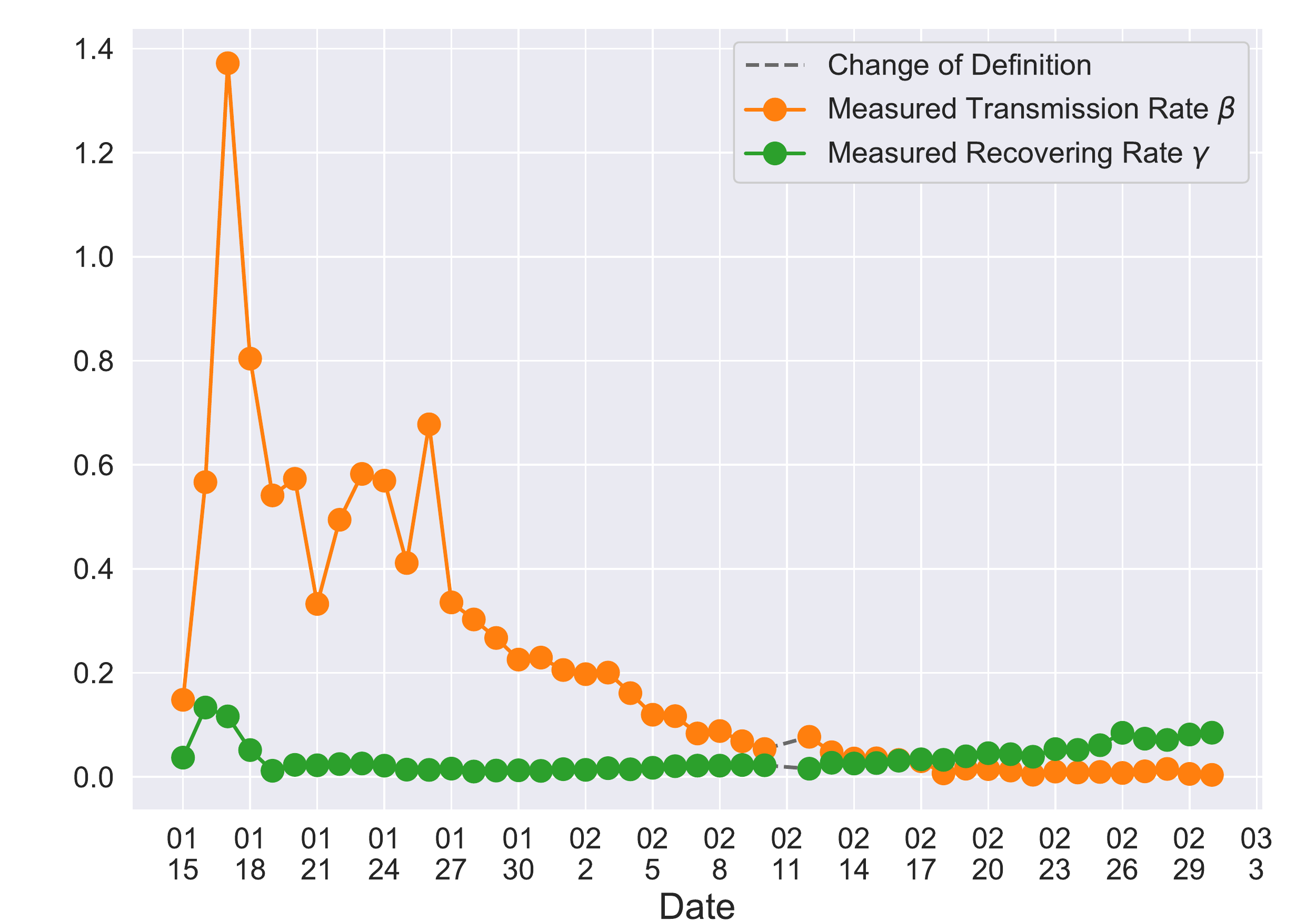}
	\caption{Measured transmission rate $\beta(t)$ and recovering rate $\gamma(t)$ of the COVID-19 from Jan. 15, 2020 to Feb. 19, 2020. The two curves are measured according to \req{d_beta} and \req{d_gamma} respectively.}
	\label{rate}
\end{figure}

In Figure \ref{rate}, we show the measured $\beta(t)$ and $\gamma(t)$ from the real historical data. We can see that $\beta(t)$ decreases dramatically, and $\gamma(t)$ increases slightly. This is a direct result of the Chinese government that tries to suppress the transmission rate $\beta(t)$ by city-wide lockdown and traffic halt. On the other hand, due to the lack of effective drugs and vaccines for COVID-19, the recovering rate $\gamma(t)$ grows relatively slowly. Additionally, there is a definition change of the confirmed case on Feb. 12, 2020 that makes the data related to Feb. 11, 2020 have no reference value. We mark these data points for $\beta(t)$ and $\gamma(t)$ with the gray dashed curve.

In an epidemic model, one crucial question is whether the disease can be contained and the epidemic will end, or whether there will be a pandemic that infects a certain fraction of the total population $n$. To answer this, one commonly used metric is the basic reproduction number $R_0$ that is defined as the average number of additional infections by an infected person before it recovers. In the classical SIR model, $R_0$ is simply $\beta/\gamma$ as an infected person takes (on average) $1/\gamma$ days to recover, and during that period time, it will be in contact with (on average) $\beta$ persons. In our time-dependent SIR model, the basic reproduction number $R_0(t)$ is a function of time, and it is defined as $\beta(t) / \gamma (t)$. If $R_0(t) > 1$, the disease will spread exponentially and infects a certain fraction of the total population $n$. On the contrary, the disease will eventually be contained. Therefore, by observing the change of $R_0(t)$ with respect to time or even predicting $R_0(t)$ in the future, we can check whether certain epidemic control policies are effective or not.

\begin{figure}[!htbp]
	\centering
	\includegraphics[width=0.48\textwidth]{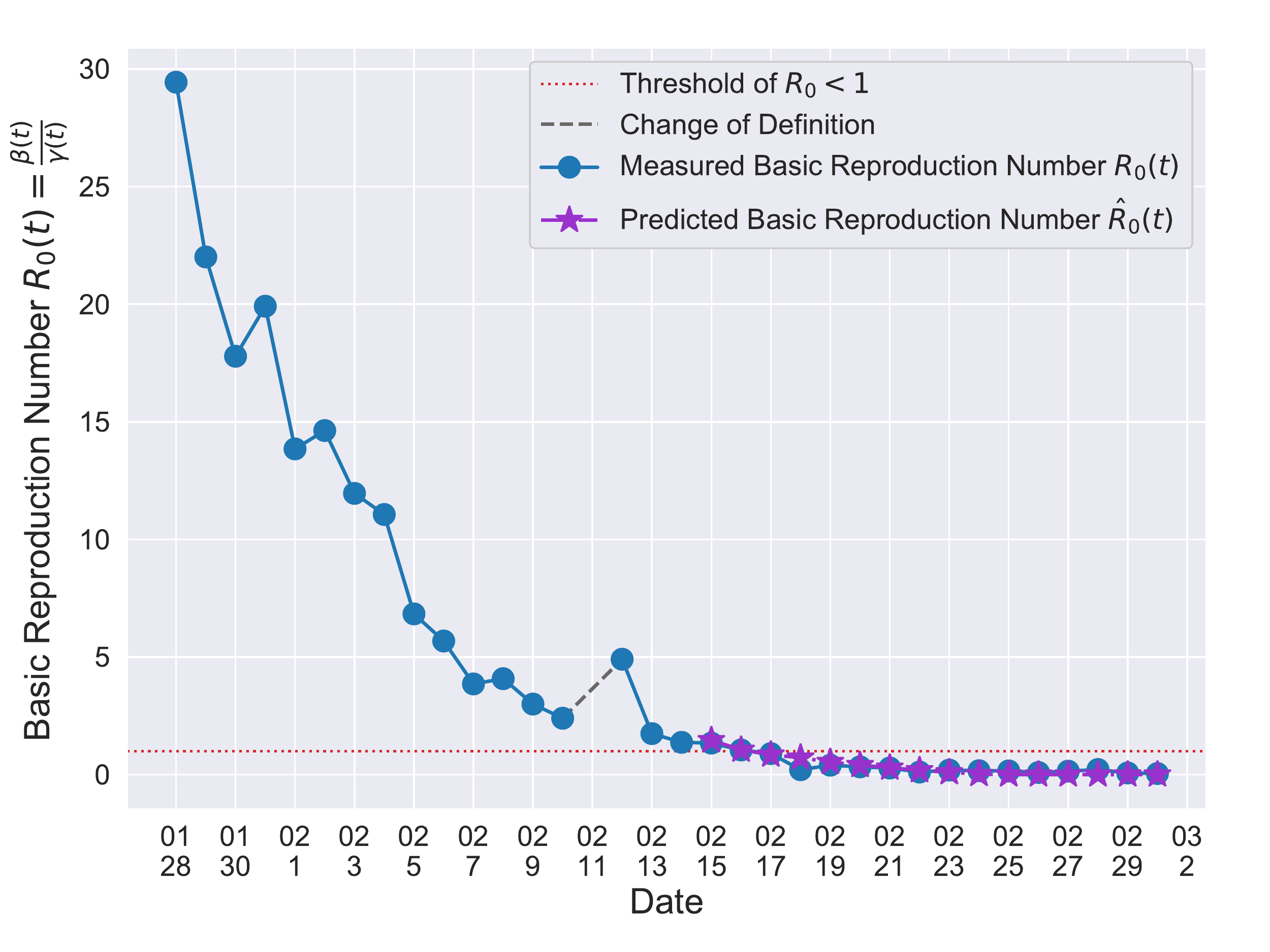}
	\caption{Basic reproduction number $R_0(t)$ of the time-dependent SIR model of the COVID-19 in China. The circle-marked solid curve with blue color is the $R_0(t)$ based on the given data from Jan. 27, 2020 to Feb. 20, 2020, the star-marked dashed curve with purple color is the predicted $\hat{R}_0(t)$ based on the data from Jan. 27, 2020 to Feb. 15, 2020, and the dashed line with red color is the percolation threshold $1$ for the basic reproduction number.}
	\label{r0_curve}
\end{figure}

In Figure \ref{r0_curve}, we show the measured basic reproduction number $R_0(t)$, and the predicted basic reproduction number $\hat{R}_0(t)$. The blue circle-marked solid curve is the measured $R_0(t)$ and the purple star-marked dashed curve is the predicted $\hat{R}_0(t)$ (from Feb. 15, 2020). It is clear that $R_0(t)$ has decreased dramatically since Jan. 28, 2020, and it implies that the control policies work in China. More importantly, it shows that the turning point is Feb. 17, 2020 when $\hat{R}_0(t)<1$. In the following days after Feb. 17, 2020, $X(t)$ will decrease exponentially, and that will lead to the end of the epidemic in China. Our model predicts precisely that $R_0(t)$ will go less than $1$ on Feb. 17, 2020 by $3$ days in advance (Feb. 14, 2020). The results show that our model is very effective in tracking the characteristics of $\beta(t)$ and $\gamma(t)$.

\subsection{One-day Prediction}
\begin{figure}[!htbp]
	\centering
	\includegraphics[width=0.48\textwidth]{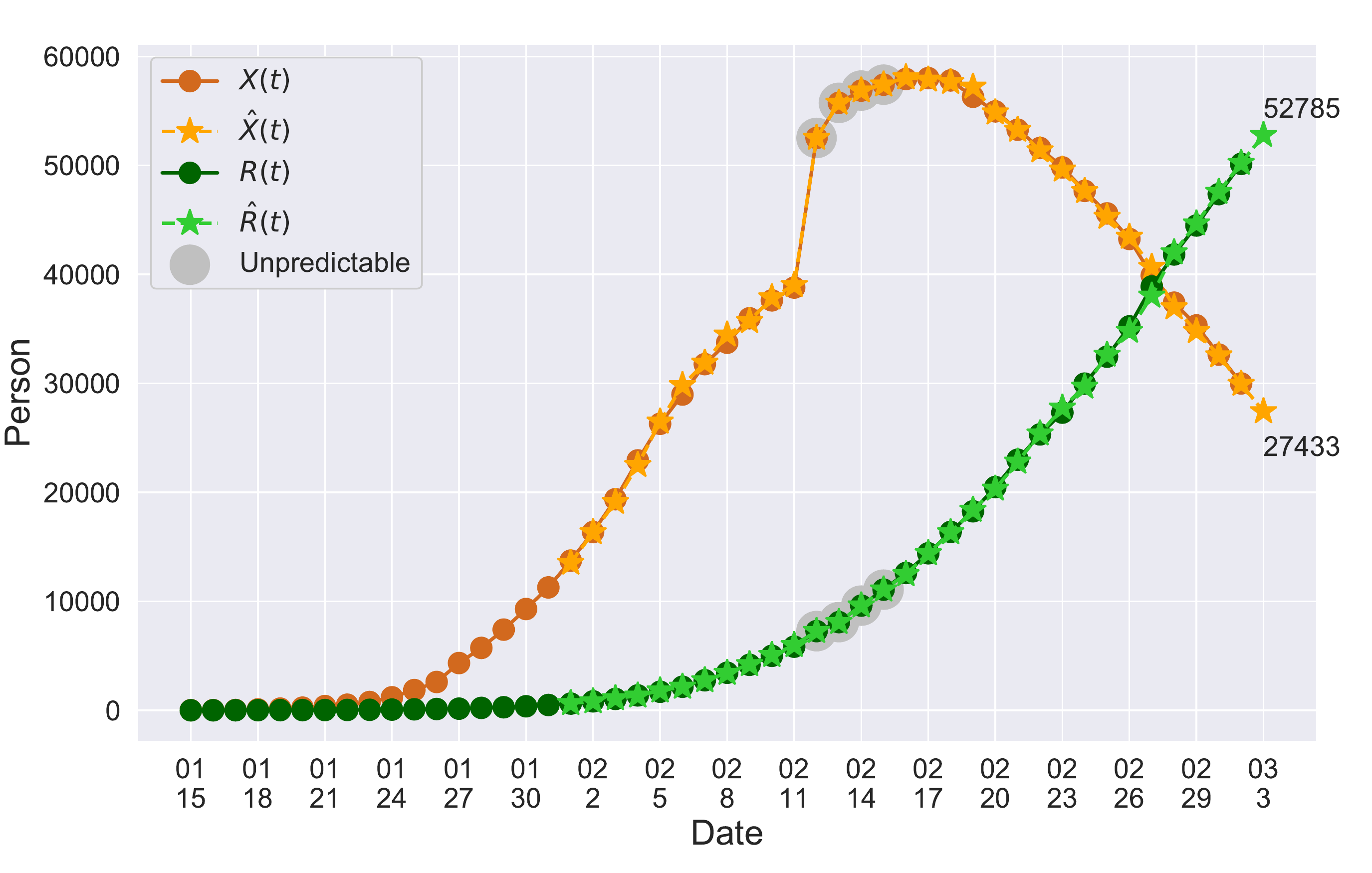}
	\caption{One-day prediction for the number of infected and recovered persons. The unpredictable points due to the change of definition of the confirmed case are marked as gray. The circle-marked solid curve with dark orange (resp. green) color is the real number of infected persons $X(t)$ (resp. recovered persons $R(t)$), the star-marked dashed curve with light orange (resp. green) color is the predicted number of infected persons $\hat{X}(t)$ (resp. recovered persons $\hat{R}(t)$).}
	\label{history_curve}
\end{figure}

To show the precision of our model, we demonstrate the prediction results for the next day (one-day prediction) in Figure \ref{history_curve}. It contains the predicted number of infected persons $\hat{X}(t)$ (orange star-marked dashed curve), the predicted number of recovered persons $\hat{R}(t)$ (green star-marked dashed curve), and the real number of infected and recovered persons (dark orange and dark green circle-marked solid curves) every day. The unpredictable days due to the change of the definition of the confirmed case on Feb. 12, 2020. are marked as gray. The predicted curves are extremely close to the measured curves (obtained from the real historical data). In this figure, we also annotate the predicted number of infected persons $\hat{X}(t)=27,433$ and the predicted number of recovered persons $\hat{R}(t)=52,785$ on Mar. 3, 2020.

\begin{figure}[!htbp]
	\centering
	\includegraphics[width=0.48\textwidth]{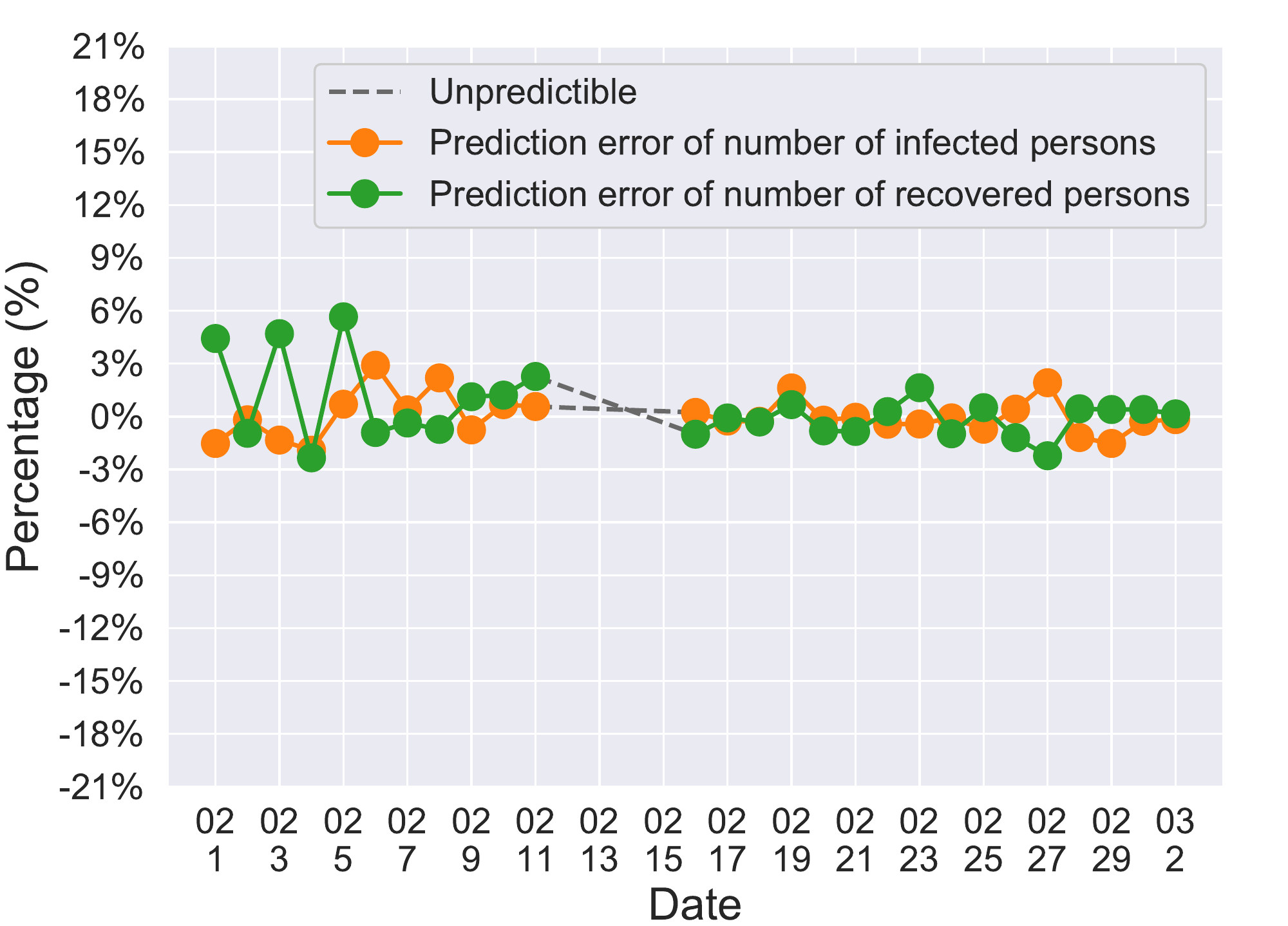}
	\caption{Errors of the one-day prediction of the number of infected and recovered persons. The unpredictable points due to the change of definition of the confirmed case on Feb. 12, 2020 are marked as the gray dash curve.}
	\label{prediction_error}
\end{figure}

We further examine our prediction accuracy in Figure \ref{prediction_error}. The error rates are all within $\pm3\%$ except for the predicted number of recovered persons $\hat{R}(t)$ on Feb. 1, Feb. 3, and Feb. 5, 2020. The gray dashed curve stands for the unpredictable points due to the change of definition of the confirmed case. However, from the prediction results after Feb. 16, 2020, we find that our model can still keep tracking $\beta(t)$ and $\gamma(t)$ accurately and overcome the impact of the change of the definition.

\subsection{Connections to the Wuhan City Lockdown}\label{history}
In this subsection, we show the connections between the epidemic prevention policies issued by the Chinese government and the historical data of the time-varying transmission rates.

\begin{figure}[!htbp]
	\centering
	\includegraphics[width=0.48\textwidth]{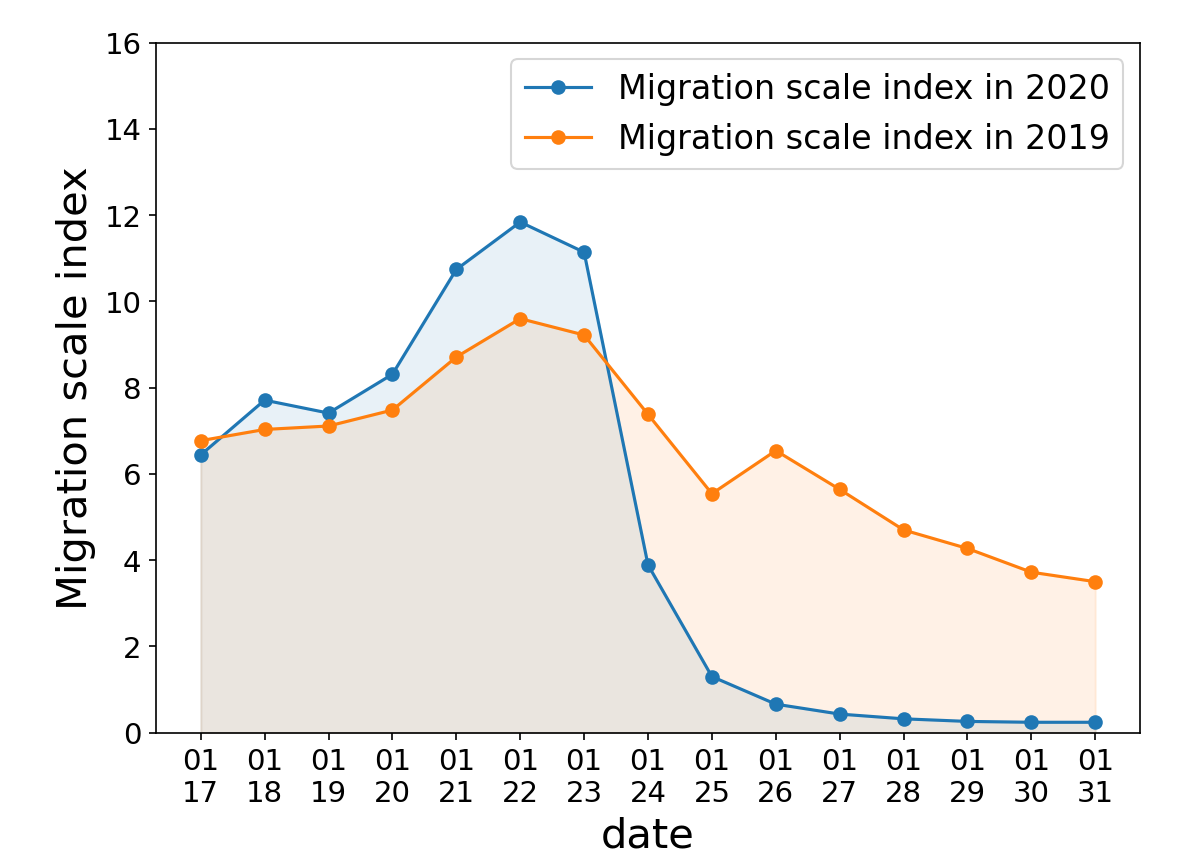}
	\caption{Trends of moving out during the Chunyun (Spring Festival travel season) in Wuhan city. The vertical axis represents the ratio between the number of people leaving the city and the resident population in Wuhan. The orange curve shows the ratio in 2019, while the blue curve shows the ratio in 2020. We redraw this figure by our-self from \url{https://qianxi.baidu.com/} \cite{baidu}.}
	\label{wuhan_outgoing}
\end{figure}

As shown in Figure \ref{rate}, the disease has been gradually controlled in China as time goes on. Excluding the small number of cases ($X(t)<500$) before Jan. 21, 2020 (that causes the curve to fluctuate a lot), it is notable that $\beta(t)$ increases gradually then drops dramatically during Jan. 23, 2020 to Jan. 28, 2020, and it reaches the peak point on Jan. 26, 2020, which coincides with the trends of the moving out in Wuhan during the Chunyun (Spring Festival travel season) \cite{baidu} in Figure \ref{wuhan_outgoing}. Especially, the emigration trend is almost the same as the $\beta(t)$ during Jan. 20, 2020 to Jan. 25, 2020. We speculate that people rushed into the public transportation system when the announcement of the Wuhan city lockdown was out, which significantly increases the contact among people and speeds up the spread of the virus. As a result of that, the transmission rate $\beta(t)$ increases substantially. As pointed out in Zhong's study \cite{guan2020clinical}, the median of the incubation period of COVID-19 is $3$ days among $1099$ valid confirmed cases, which makes the emigration trend aligns with $\beta(t)$ if the extra $3$ days are taken into account. Finally, the disease is gradually getting under control after the lockdown. The basic reproduction number $R_0(t)$ is less than $1$, i.e., $\beta(t) < \gamma(t)$ since Feb. 17, 2020.

\subsection{Basic Reproduction Numbers of Several Other Countries}\label{country}
In addition to the dataset for China, we also measure the basic reproduction number $R_0(t)$ on Mar. 31, 2020 for several countries from the datasets in \cite{cssegisanddata_johns}. This is shown in the last column of Table \ref{tab:country_r0}. As the data for the cumulative numbers of recovered persons for these countries are noisy, we also show the estimated $R_0(t)$ under various assumptions of the average time to recover $1/\gamma$. The $R_0(t)$ values for the five countries, including United States of America, the United Kingdom, France, Iran, and Spain are very high. On the other hand, it seems that Italy is gaining control of the spread of the disease after the Italian government announces the lockdown and forbids the gatherings of people on Mar. 10, 2020. Also, both Germany and Republic of Korea are capable of controlling the spread of the disease.
\begin{table*}[!htbp]
	\centering
	\resizebox{\textwidth}{!}{%
		\begin{tabular}{|c|c|c|c|c|c|c|}
			\hline
			\multirow{2}{*}{Country} &
			\multicolumn{5}{c|}{Estimated $R_0(t)$ when the average time to recover $1/\gamma$ is} &
			\multirow{2}{*}{\begin{tabular}[c]{@{}c@{}}$R_0(t)$ on\\ Mar. 31, 2020\end{tabular}}\\ \cline{2-6}
			&
			$14$ Days &
			$21$ Days &
			$28$ Days &
			$35$ Days &
			$42$ Days &
			\\ \hline
			United States of America & 2.13 & 3.20 & 4.26 & 5.33 & 6.39 & 12.59 \\ \hline
			The United Kingdom       & 1.89 & 2.83 & 3.77 & 4.72 & 5.66 & 8.90  \\ \hline
			France                   & 2.58 & 3.86 & 5.15 & 6.44 & 7.73 & 4.76  \\ \hline
			Iran                     & 1.25 & 1.88 & 2.51 & 3.14 & 3.76 & 4.51  \\ \hline
			Spain                    & 1.63 & 2.45 & 3.27 & 4.08 & 4.90 & 3.47  \\ \hline
			Italy                    & 0.79 & 1.19 & 1.59 & 1.98 & 2.38 & 3.08  \\ \hline
			Germany                  & 1.49 & 2.24 & 2.98 & 3.73 & 4.48 & 2.80  \\ \hline
			Republic of Korea        & 0.44 & 0.66 & 0.88 & 1.11 & 1.33 & 1.68  \\ \hline
		\end{tabular}%
	}
	\caption{The estimated $R_0(t)$ under various assumptions of the average time to recover ($1/\gamma$) from COVID-19, and the measured $R_0(t)$ on Mar. 31, 2020.}
	\label{tab:country_r0}
\end{table*}

\subsection{The Effects of Type II Infected Persons}\label{perco_experiment}
In this subsection, we show how undetectable (type \RomanNumeralCaps{2}) infected persons affect the epidemic. In particular, we are interested in addressing the question of whether the existence of undetectable infected persons (type \RomanNumeralCaps{2}) can cause an outbreak.

To carry out our numerical study, we need to fix some variables in the system of difference equations in (\ref{diff_sir_undetect1})-(\ref{diff_sir_undetect3}). For the transmission rate (resp. recovering rate) of type \RomanNumeralCaps{1} infected persons, i.e., $\beta_1$ (resp. $\gamma_1$), we set it to be the measured $\beta(t)=0.00383$ (resp. $\gamma(t)=0.08493$) on Mar. 1, 2020 in China. The rationale behind this is that type \RomanNumeralCaps{1} infected persons were detected and they were under treatment and isolation after Mar. 1, 2020 in China. Also, as there is no medicine for COVID-19, we may assume that these two types of infected persons have the same recovering rate, i.e, $\gamma_2=\gamma_1$. In view of the system equation in (\ref{diff_sir_undetect1})-(\ref{diff_sir_undetect3}), there are still two free variables $w_2$ and $\beta_2$, where $w_2$ is the probability that an infected person is of type \RomanNumeralCaps{2} and $\beta_2$ is the transmission rate of type \RomanNumeralCaps{2}.

\begin{figure}[!htbp]
	\centering
	\includegraphics[width=0.48\textwidth]{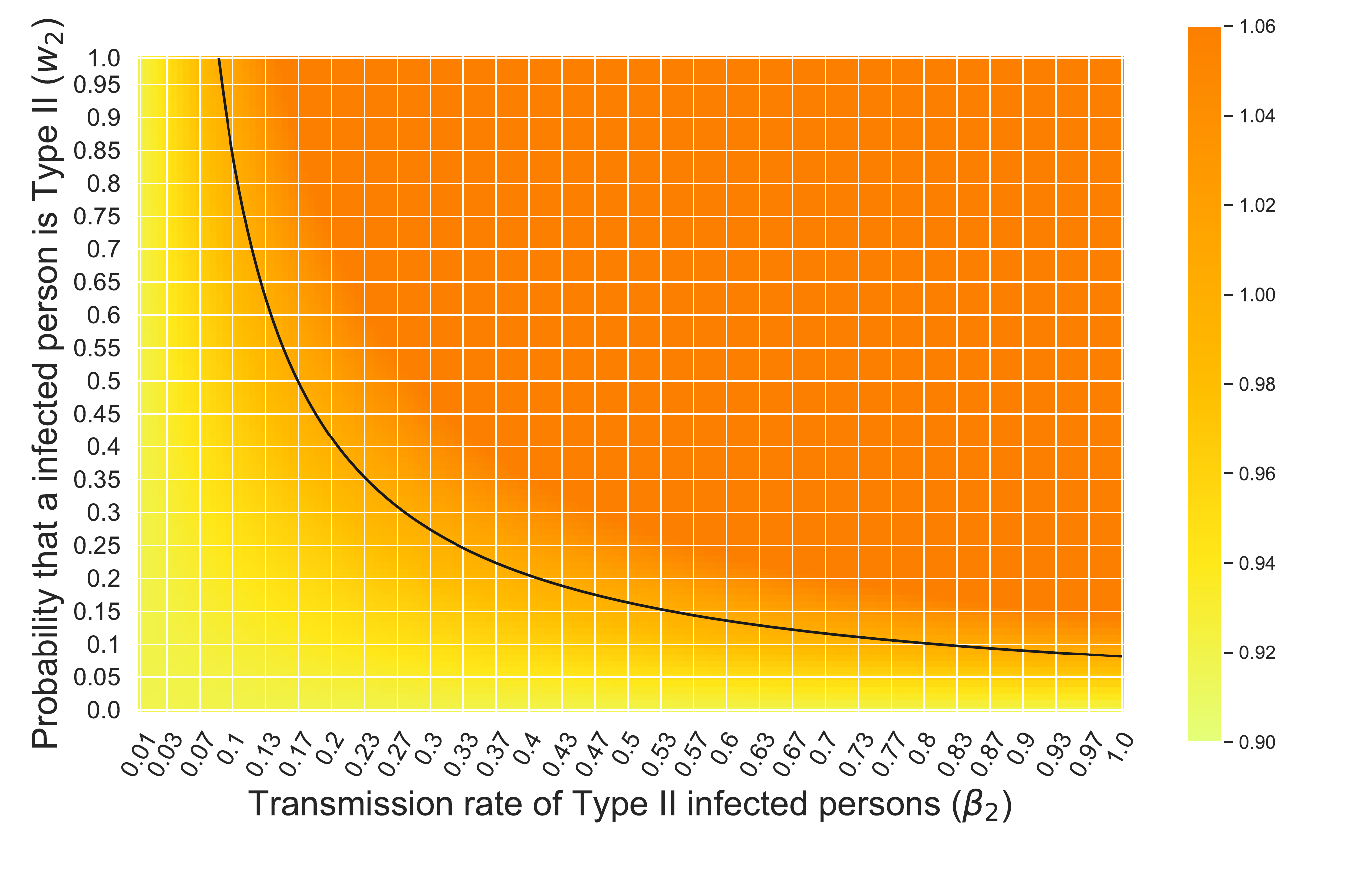}
	\caption{Phase transition diagram of an outbreak with respect to $\beta_2$ and $w_2$. The black curve is the percolation threshold. The orange area means the disease will be an outbreak while the yellow area means the disease is under control.}
	\label{perco_threshold_p2b2}
\end{figure}

In Figure \ref{perco_threshold_p2b2}, we illustrate how $w_2$ and $\beta_2$ affect the outbreak of the COVID-19. Such a figure is known as the phase transition diagram in \cite{newman2010networks}. The black curve in Figure \ref{perco_threshold_p2b2} is the curve when the spectral radius of the transition matrix $\bf A$ in (\ref{eq:transition_matrix}) equals to $1$. This curve represents the percolation threshold of COVID-19. If $w_2$ and $\beta_2$ fall above the black curve (in the orange zone), then there will be an outbreak. On the contrary, if $w_2$ and $\beta_2$ fall below the black curve (in the yellow zone), then there will not be an outbreak. As shown in Figure \ref{perco_threshold_p2b2}, we would like to point out the importance of detecting an infected person. As long as more than $90\%$ of those infected persons can be actually detected and properly isolated and treated, it is possible to contain the spread of the disease even if the transmission rate of type \RomanNumeralCaps{2} infected persons, i.e., $\beta_2$, is as high as $0.7$. On the other hand, suppressing the transmission rate of type \RomanNumeralCaps{2} infected persons can also be effective in controlling the disease while the detection rate is not that high. For example, wearing masks and washing hands can be an effective epidemic prevention mechanism to reduce $\beta_2$.

In the following experiments, we extend our study to other countries, including Japan, Singapore, South Korea, Italy, and Iran. We collect the historical data from Jan. 22, 2020 to Mar. 2, 2020 from the GitHub of Johns Hopkins University \cite{cssegisanddata_johns}. For these countries, the transmission rates $\beta(t)$ and the recovering rates $\gamma(t)$ (measured from the time-dependent SIR model in Section \ref{td_sm}) during the initial period with a rapid increase of the number of confirmed cases can be viewed as $\beta_2$ and $\gamma_2$. This is because during that period of time, there is no epidemic prevention intervention, and all the infected persons are basically not detected. It is interesting to note that different countries might have different $\beta_2$ and $\gamma_2$. On the other hand, only $87.9\%$ of COVID-19 cases have a fever from the report of WHO \cite{who}. If we use body temperature as a means to detect an infected person, then only $87.9\%$ of COVID-19 cases can be detected. For this, we set $w_1=87.9\%$.

\begin{figure}[!htbp]
	\centering
	\includegraphics[width=0.48\textwidth]{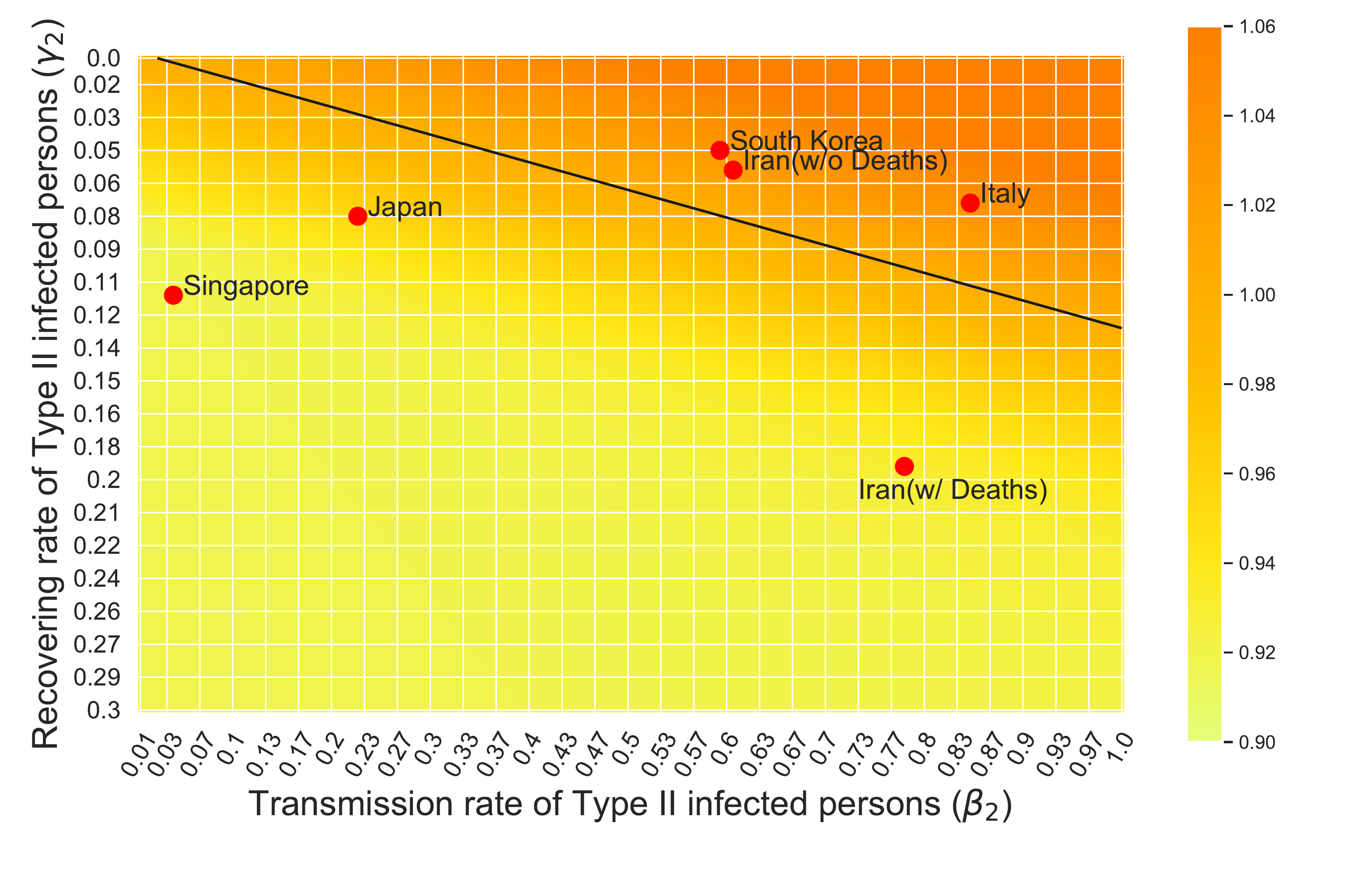}
	\caption{Phase transition diagram of an outbreak with respect to $\beta_2$ and $\gamma_2$. The black curve is the percolation threshold. The orange area means the disease will be an outbreak, while the yellow area means the disease is under control.}
	\label{perco_threshold_r2b2}
\end{figure}

With $\beta_1$ and $\gamma_1$ specified in the previous study for China, we plot the phase transition diagram in Figure \ref{perco_threshold_r2b2} in terms of the two variables $\beta_2$ and $\gamma_2$. Again, the black curve is the curve when the spectral radius of the transition matrix $\bf A$ in (\ref{eq:transition_matrix}) equals to $1$. Such a curve represents the percolation threshold of a COVID-19 outbreak. If $\beta_2$ and $\gamma_2$ fall above the black curve (in the orange zone), then there will be an outbreak. On the contrary, if $\beta_2$ and $\gamma_2$ fall below the black curve (in the yellow zone), then there will not be an outbreak. The countries with large confirmed cases, including Japan, Singapore, South Korea, Italy, and Iran, are marked in Figure \ref{perco_threshold_r2b2}. From Figure \ref{perco_threshold_r2b2}, we observe that both Singapore and Japan are below the percolation threshold. But Japan is much closer to the percolation threshold. On the other hand, both South Korea and Italy are above the percolation threshold, and they are on the verge of a potential outbreak on Mar. 2, 2020. These two countries must implement epidemic prevention policies urgently. Not surprisingly, on Mar. 10, 2020, the Italian government announces the lockdown and forbids the gatherings of people. It is worth mentioning that there are two marks for Iran in the Figure. The one above the percolation threshold is the one without adding the number of deaths into the number of recovered persons. The other one below the percolation threshold is the one that adds the number of deaths into the number of recovered persons. For some unknown reason, the death rate in Iran is higher than the other countries. The high death rate seems to prevent an outbreak in Iran.

\subsection{The Effects of Social Distancing}
\begin{figure}[!htbp]
	\centering
	\includegraphics[width=0.48\textwidth]{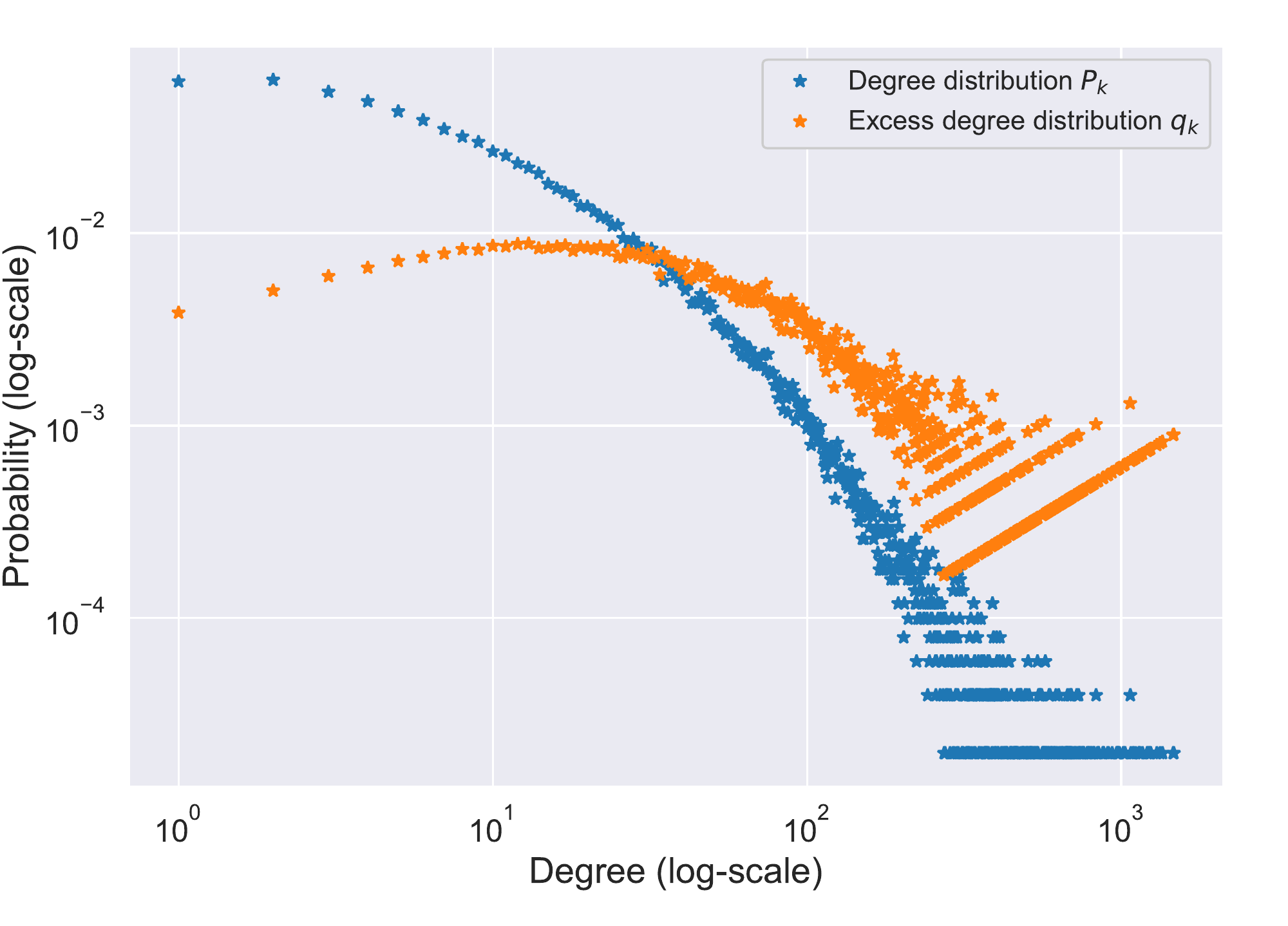}
	\caption{The degree distribution and the excess degree distribution of the Facebook dataset.}
	\label{facebook}
\end{figure}

In this subsection, we show the numerical results of the social distancing approach that cancels mass gatherings by removing nodes with the number of edges larger than or equal to $k_0$. As shown in \rthe{socialdisb}, the basic reproduction number is reduced by a factor of $\displaystyle\frac{\sum_{k=0}^{k_0-2}k q_k}{\sum_{k=0}^{\infty}k q_k}$, where $q_k$ is the excess degree distribution of $p_k$. For this experiment, we use the dataset collected by \cite{rozemberczki2019gemsec} from Facebook. This dataset represents the verified Facebook page (with blue checkmark) networks of the artist category. The blue checkmark means Facebook has confirmed that an account is the authentic presence of the public figure, celebrity, or global brand it represents. Each node in the network represents the page, and edges between two nodes are mutual likes among them. This dataset is composed of $50,515$ nodes and $819,306$ edges. Some other properties are listed as follow: mean degree $32.4$, max degree $1,469$, diameter $11$, and clustering coefficient $0.053$. In Figure \ref{facebook}, we show the log-log plots of the degree distribution and the excess degree distribution of this dataset. The degree distribution appears to be a (truncated) Pareto distribution with the exponent $1.69$ (the slope in the figure).

\begin{figure}[!htbp]
	\centering
	\includegraphics[width=0.48\textwidth]{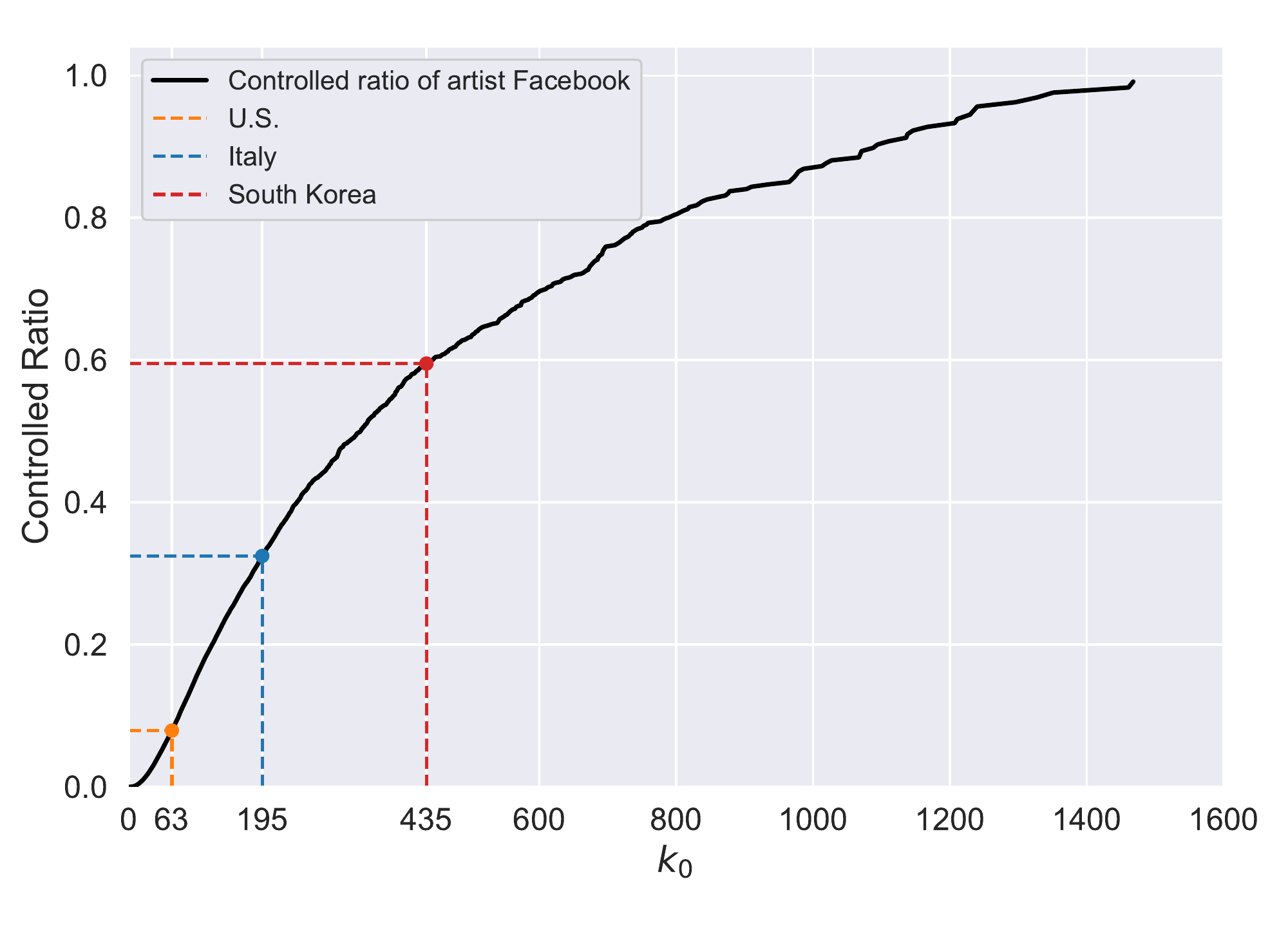}
	\caption{The reduction ratio $\frac{\sum_{k=0}^{k_0-2}k q_k}{\sum_{k=0}^{\infty}k q_k}$ as a function of $k_0$. The minimum $k_0$s to prevent an outbreak in Italy, U.S., and South Korea are $63$, $195$, and $435$, respectively.}
	\label{facebookk0}
\end{figure}

In Figure \ref{facebookk0}, we plot the reduction ratio $\frac{\sum_{k=0}^{k_0-2}k q_k}{\sum_{k=0}^{\infty}k q_k}$ as a function of $k_0$. The ratio is between $0$ and $1$, and it is monotonically increasing in $k_0$. Using the $R_0$ values (on Mar. 31, 2020) in the last column of Table \ref{tab:country_r0}, we also show that the minimum $k_0$s to prevent an outbreak in Italy, U.S., and South Korea are $63$, $195$, and $435$, respectively; moreover, the affected fraction of tail distributions are $13.1\%$, $2.2\%$, and $0.4\%$, respectively. In particular, if canceling mass gathering is the only measure used for controlling COVID-19 in the U.S. with the $R_0$ value of $12.59$ on Mar. 31, 2020, then one can prevent an outbreak by ``removing'' all the nodes with degrees larger than or equal to $63$ (in the Facebook dataset), and the removal might affect $13.1\%$ of the nodes in the Facebook dataset.

\subsection{The Ratio of the Population Infected in the Long Run}

\begin{figure}[!htbp]
	\centering
	\includegraphics[width=0.48\textwidth]{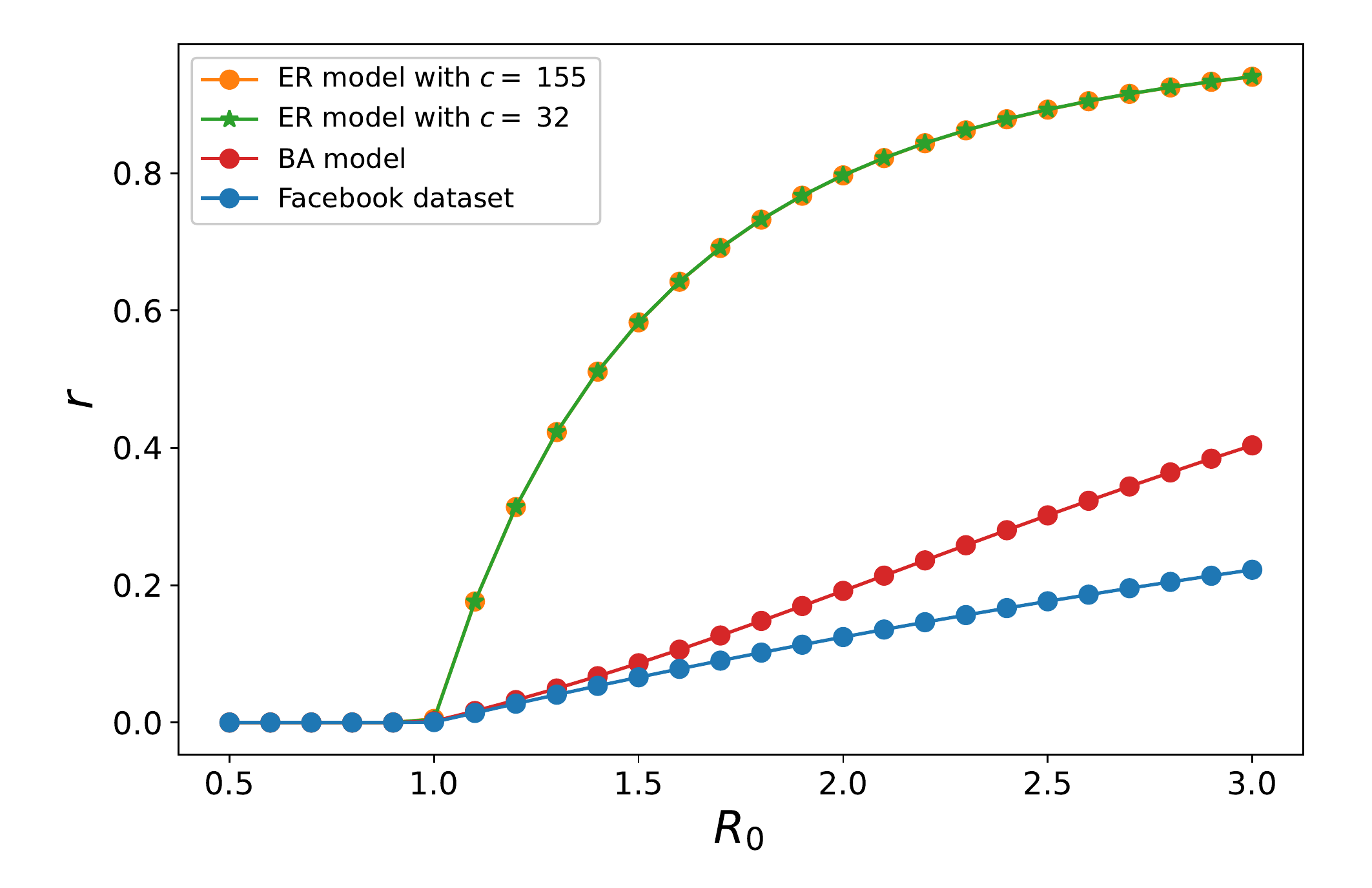}
	\caption{The probability $r$ that a randomly selected node is infected in the long run as a function of $R_0$ for various degree distributions.}
	\label{S_and_r0}
\end{figure}

In this subsection, we show the numerical results of the ratio of the population infected in the long run in \rsubsec{longrun}. As shown in \rthe{longrun}, if $R_0 >1$, then there is a nonzero probability $r$ that a randomly selected node is infected in the long run. In Figure \ref{S_and_r0}, we plot the infected probability $r$ as a function of $R_0$ for various degree distributions. For this numerical experiment, we use the same Facebook dataset in the previous subsection. Note that the mean degree and the mean excess degree of this Facebook dataset are $32.4$ and $155.6$, respectively. We also generate three random networks: two from the Erd\"os-R\'enyi (ER) model \cite{erdos1959random} with the mean degrees $c=32$ and $c=155$, and one from the Barab\'asi-Albert (BA) model \cite{barabasi1999emergence} with the mean degree $c=32$. The numbers of nodes are all set to $50,000$. The degree distribution of the BA model (generated by using the linear preferential attachment rule) is known to follow the asymptotic power-law distribution with $p_k\approx k^{-3}$ for large $k$'s. The mean excess degree of the BA model is $84.05$.

From Figure \ref{S_and_r0}, one can see that there is an outbreak when $R_0 > 1$. The probability $r$ is increasing in $R_0$. Also, as shown in this figure, the two curves of the two ER models (even with different mean degrees) overlap with each other. That means that $r$ is independent of the mean degree when the degree distribution is Poisson (as shown in \req{Poisson1234} at the end of \rsubsec{longrun}).

On interesting observation is that the infected probability $r$ of the ER model is larger than the infected probabilities of the other networks with the power-law degree distributions, i.e., the Facebook dataset and the BA model. This is because we relate the propagation probability $\phi$ by $R_0/g_1^\prime(1)$ in \req{long4444} (to ensure that the average number of additional infections by an infected person in the IC model is the same as that in the SIR model). Thus, a network with a large mean excess degree $g_1^\prime(1)$ has a low propagation probability $\phi$ (when $R_0$ is fixed). That leads to a smaller ratio of the population infected in the long run. To explain this further, we note that the probability of having a super spreader (a node with a very large degree) in a network with a power-law degree distribution is higher than that of the ER model. A super spreader is capable of infecting a large number of its first neighbors. However, the neighbors of a super spreader tend to have relatively low degrees (contacts) than that of a super spreader. As such, it makes the disease more difficult to spread further in the IC model. On the other hand, the probability of having a super spreader in the ER model is exponentially small. Since the infected subgraph appears to be a tree (if one follows an edge of an infected node to propagate the disease to the other nodes in the IC model), the disease may spread slowly in the early stage of the epidemic in the ER model than that in the BA model. However, if $R_0>1$, the disease may continue to propagate in the ER model to a larger fraction of the total population in the long run. For example, when $R_0=3$, the percentage of infected persons is more than $90\%$ in the ER model.

\section{Discussions and Suggestions}\label{discussion}
As of Mar. 2, 2020, it seems that COVID-19 has been gradually controlled in China since the prevention policies, such as city-wide lockdown was issued in China in Jan. 2020. Although the policies such as traffic halt, small community management, and city-wide lockdown can effectively reduce the transmission rates $\beta_1$ and $\beta_2$; however, these relatively extreme policies not only restrict the right of personal freedom but also affect the normal operation of society. These extreme policies forced several companies and factories to halt production, which impacts all sectors of the economy. Therefore, to strike a balance between the prevention of disease and ensuring the normal operation of society is crucial, and it is important to suggest the so-called ``optimal'' control policies.

For this, we would like to put forward some discussions and suggestions for controlling the spread of COVID-19 based on the observation made from the results of our system models. From our results in \rthe{stability}, \rcor{herd}, \rthe{stabilityIC}, \rcor{socialdis}, and \rthe{socialdisb}, we know that there is no outbreak for a disease if
\beq{summary1111}
h \cdot s \cdot (w_1 \frac{\beta_1}{\gamma_1} +w_2 \frac{\beta_2}{\gamma_2})<1,
\eeq
where $h$ is the herd immunity reduction factor in \rcor{herd} when $1-h$ fraction of individuals are immune to the disease, $s$ is the social distancing reduction factor in \rcor{socialdis} or \rthe{socialdisb}, $w_1$ (resp. $w_2$) is the probability that an infected person can (resp. cannot) be detected, and $\beta_1$ and $\gamma_1$ (resp. $\beta_2$ and $\gamma_2$) are the transmission rate and the recovering rate of an infected person who can (resp. cannot) be detected. For COVID-19, $\beta_1 \le \beta_2$ and $\gamma_1 \ge \gamma_2$ as an infected person, once detected, can be treated (to shorten the recovery time) and isolated (to reduce the transmission rate). As $w_1+w_2=1$, it is thus preferable to have a much larger $w_1$ than $w_2$.

To prevent an outbreak, one should minimize the value on the left-hand side of \req{summary1111}. Here we discuss several approaches for that.

\begin{itemize}
	\item [1.] Increasing the recovering rate $\gamma_1$: the most effective way to increase $\gamma_1$ is to find anti-virus drugs; however, it takes time. Hence, we should focus on the other approaches to control the spread of the disease in this stage.
	
	\item[2.] Reducing the herd immunity reduction factor $h$: the most effective way for this is to find vaccines to reduce the fraction of susceptible persons. Once again, it takes time, and we should focus on the other approaches to control the spread of the disease in this stage.
	
	\item[3.] Decreasing the transmission rate $\beta_1$: once an infected person is detected, it should be isolated to avoid extra infection on society and lower the transmission rate $\beta_1$. Quarantine of the persons who are suspected to be in contact with infected persons could also lower the transmission rates $\beta_1$ and $\beta_2$.
	
	\item [4.] Increasing the detection probability $w_1$ (and thus reducing $w_2$): mass testing can certainly increase $w_1$. In fact, South Korea did an outstanding job of drive-thru testing. If mass testing is not possible due to the limitation of medical resources, then measuring body temperature can also be an effective alternative, as $87.9\%$ of the confirmed cases of COVID-19 have a fever. In addition to this, one can also track the travel history, occupation, contact, and cluster (TOCC) of the confirmed cases to narrow the range of the possible sources. These sources might contain asymptomatic infected persons, and testing the close contacts of these possible sources can thus increase $w_1$ by detecting asymptomatic infected persons.
	
	\item [5.] Decreasing the transmission rate $\beta_2$: propaganda of health education knowledge can reduce the transmission rate $\beta_2$ substantially. For example, wearing masks in public and enclosed space, washing hands, avoiding touching your mouth, eyes, and nose are good ways to not only protect ourselves from being infected by the asymptomatic infected persons but also avoid infecting others.
	
	\item[6.] Reducing the social distancing reduction factor $s$: as shown in \rcor{socialdis} and \rthe{socialdisb}, there are two practical approaches that can reduce the social distancing reduction factor $s$: (i) allowing every person to keep its interpersonal contacts up to a fraction of its normal contacts, and (ii) canceling mass gatherings.
\end{itemize}

\section{Conclusion and Future Work}\label{conclude}
In this paper, we conducted mathematical and numerical analyses for COVID-19. Our time-dependent SIR model is not only more adaptive than traditional static SIR models, but also more robust than direct estimation methods. Our numerical results show that one-day prediction errors for the number of infected persons $X(t)$ and the number of recovered persons $R(t)$ are within (almost) $3\%$ for the dataset collected from the National Health Commission of the People’s Republic of China (NHC) \cite{outbreak_notification_2020}. Moreover, we are capable of tracking the characteristics of the transmission rate $\beta(t)$ and the recovering rate $\gamma(t)$ with respect to time $t$, and precisely predict the future trend of the COVID-19 outbreak in China.

To address the impact of asymptomatic infections in COVID-19, we extended our SIR model by considering two types of infected persons: detectable infected persons and undetectable infected persons. Whether there is an outbreak in such a model is characterized by the spectral radius of a $2 \times 2$ matrix that is closely related to the basic reproduction number $R_0$. In addition to our numerical analysis for China, we further extended our study to other countries, including Japan, Singapore, South Korea, Italy, and Iran.

To understand the effects of social distancing approaches, including the reduction of interpersonal contacts and canceling mass gatherings, we analyzed the IC model for disease propagation in the configuration model. By relating the propagation probabilities in the IC model to the transmission rates and recovering rates in the SIR model, we showed these social distancing approaches can lead to a reduction of $R_0$.

Last but not least, based on the experimental results, some discussions and suggestions on epidemic prevention are proposed from the perspective of our models. In the future, we would like to extend our deterministic SIR model by using stochastic models, such as the non-homogeneous Markov chain, to further improve the precision of the prediction results.

\ifCLASSOPTIONcaptionsoff
\newpage
\fi



\bibliographystyle{IEEEtran}
\bibliography{COVID-19_SIR_arXiv_0428}
%

%

\begin{IEEEbiography}[{\includegraphics[width=1in,height=1.25in,clip,keepaspectratio]{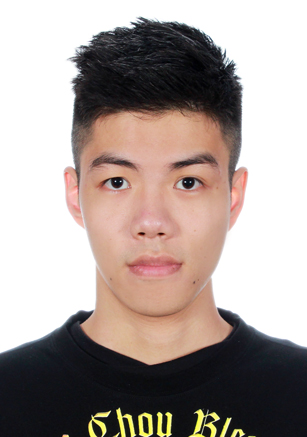}}]{Yi-Cheng Chen}
	received his B.S. degree in electrical engineering from National Taiwan University of Science and Technology, Taipei, Taiwan (R.O.C.), in 2018. He is currently pursuing the M.S. degree in the Institute of Communications engineering, National Tsing-Hua University, Hsinchu, Taiwan (R.O.C.).
\end{IEEEbiography}

\begin{IEEEbiography}[{\includegraphics[width=1in,height=1.25in,clip,keepaspectratio]{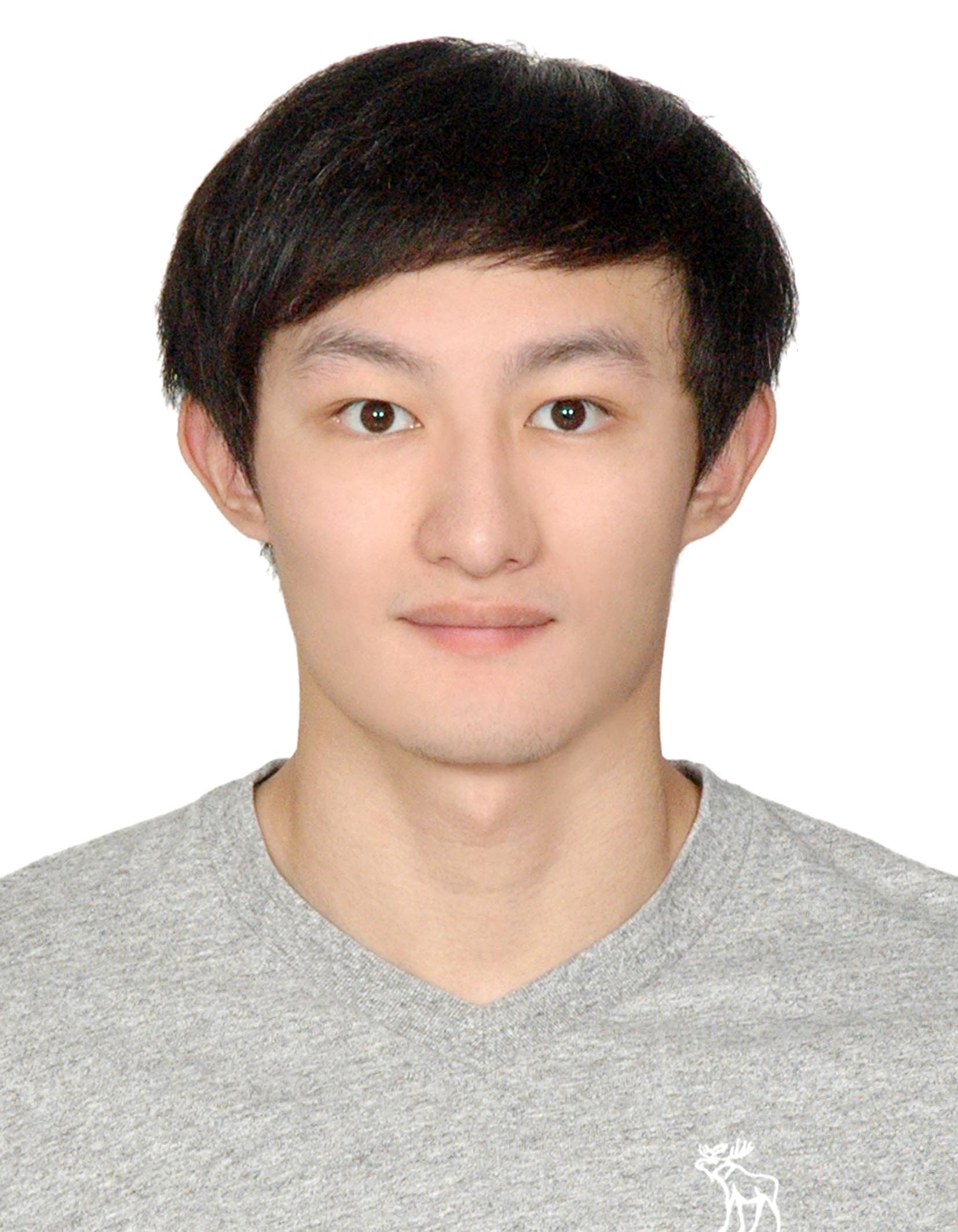}}]{Ping-En Lu} (GS'17)
	received his B.S. degree in communication engineering from the Yuan Ze University, Taoyuan, Taiwan (R.O.C.), in 2015. He is currently pursuing the Ph.D. degree in the Institute of Communications Engineering, National Tsing Hua University, Hsinchu, Taiwan (R.O.C.). He won the ACM Multimedia 2017 Social Media Prediction (SMP) Challenge with his team in 2017. His research interest is in network science, efficient clustering algorithms, network embedding, and deep learning algorithms. He is an IEEE Graduate Student Member.
\end{IEEEbiography}


\begin{IEEEbiography}[{\includegraphics[width=1in,height=1.25in,clip,keepaspectratio]{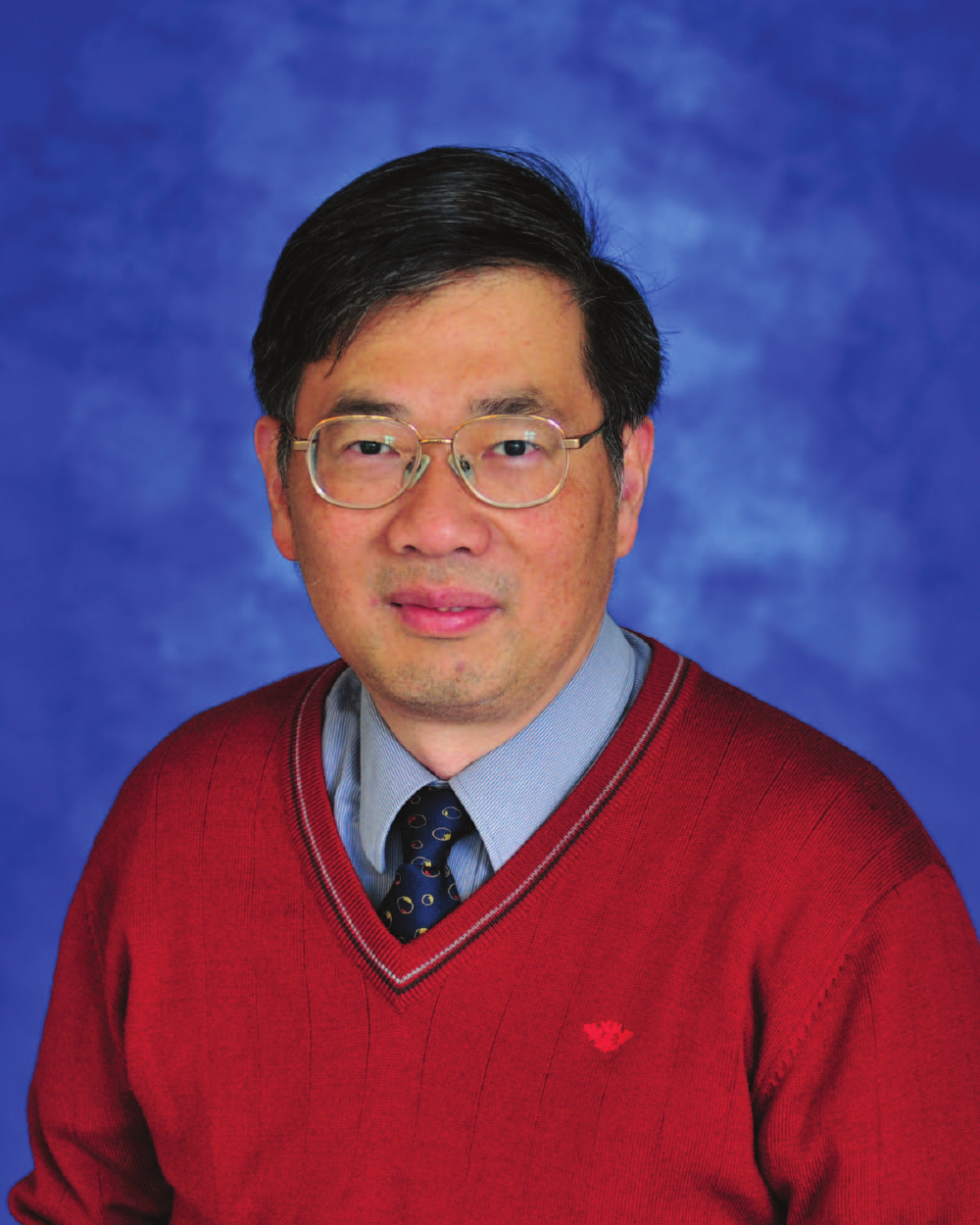}}]{Cheng-Shang Chang} (S'85-M'86-M'89-SM'93-F'04)
	received the B.S. degree from National Taiwan University, Taipei, Taiwan, in 1983, and the M.S. and Ph.D. degrees from Columbia University, New York, NY, USA, in 1986 and 1989, respectively, all in electrical engineering.
	
	From 1989 to 1993, he was employed as a Research Staff Member with the IBM Thomas J. Watson Research Center, Yorktown Heights, NY, USA. Since 1993, he has been with the Department of Electrical Engineering, National Tsing Hua University, Taiwan, where he is a Tsing Hua Distinguished Chair Professor. He is the author of the book Performance Guarantees in Communication Networks (Springer, 2000) and the coauthor of the book Principles, Architectures and Mathematical Theory of High Performance Packet Switches (Ministry of Education, R.O.C., 2006). His current research interests are concerned with network science, big data analytics, mathematical modeling of the Internet, and high-speed switching.
	
	Dr. Chang served as an Editor for Operations Research from 1992 to 1999, an Editor for the {\em IEEE/ACM TRANSACTIONS ON NETWORKING} from 2007 to 2009, and an Editor for the {\em IEEE TRANSACTIONS ON NETWORK SCIENCE AND ENGINEERING} from 2014 to 2017. He is currently serving as an Editor-at-Large for the {\em IEEE/ACM TRANSACTIONS ON NETWORKING}. He is a member of IFIP Working Group 7.3. He received an IBM Outstanding Innovation Award in 1992, an IBM Faculty Partnership Award in 2001, and Outstanding Research Awards from the National Science Council, Taiwan, in 1998, 2000, and 2002, respectively. He also received Outstanding Teaching Awards from both the College of EECS and the university itself in 2003. He was appointed as the first Y. Z. Hsu Scientific Chair Professor in 2002. He received the Merit NSC Research Fellow Award from the National Science Council, R.O.C. in 2011. He also received the Academic Award in 2011 and the National Chair Professorship in 2017 from the Ministry of Education, R.O.C. He is the recipient of the 2017 IEEE INFOCOM Achievement Award.
\end{IEEEbiography}

\begin{IEEEbiography}[{\includegraphics[width=1in,height=1.25in,clip,keepaspectratio]{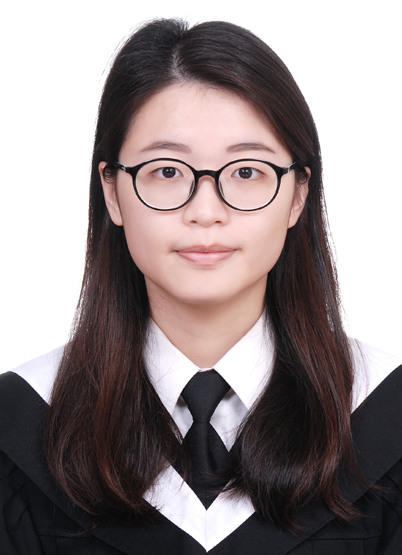}}]{Tzu-Hsuan Liu}
	received the B.S. degree in communication engineering from National Central University, Taoyuan, Taiwan (R.O.C.), in 2018. She is currently pursuing the M.S. degree in the Institute of Communications Engineering, National Tsing Hua University, Hsinchu, Taiwan (R.O.C.). Her research interest is in 5G wireless communication.
\end{IEEEbiography}




\end{document}